\documentclass[smallextended]{svjour3}       % onecolumn (second format)
\smartqed  % flush right qed marks, e.g. at end of proof
\usepackage{graphicx}
\usepackage{hyperref}
\journalname{Bulletin of Mathematical Biology}
\begin{document}

\title{Modelling the ATP production in mitochondria}


\author{Alberto Saa         \and
        Kellen M. Siqueira
}


\institute{Alberto Saa \at
Departamento de Matem\'atica Aplicada, \\
Universidade Estadual de Campinas,\\
13083-859 Campinas,  SP, Brazil \\
              \email{asaa@ime.unicamp.br}           %  \\
           \and
           Kellen M. Siqueira \at
Instituto de F\'\i sica ``Gleb Wataghin'',\\
Universidade Estadual de Campinas, \\
13083-859 Campinas,  SP, Brazil \\
\email{kellenms@ifi.unicamp.br}
}

\date{Received: date / Accepted: date}

\maketitle

\begin{abstract}
We revisit here the mathematical model for ATP production in mitochondria introduced recently by Bertram, Pedersen, Luciani, and Sherman (BPLS) as a simplification of the more complete but intricate Magnus and Keizer's model. 
We identify some inaccuracies  in the BPLS 
original approximations for two flux rates, namely the 
adenine nucleotide translocator  rate 
$J_{\rm ANT}$ and  the calcium uniporter  rate 
$J_{\rm uni}$.
We introduce new approximations for such flux rates 
and then analyze some of the dynamical properties of the model. We infer, from exhaustive numerical explorations, that 
the enhanced BPLS equations have a unique attractor fixed point
for physiologically acceptable  ranges of mitochondrial variables and respiration inputs,
as one would indeed expect from homeostasis.
We determine, in the stationary regime, the dependence of the   mitochondrial variables
on the respiration inputs, namely the cytosolic concentration of calcium ${\rm Ca}_{\rm c}$ and the substrate fructose 1,6-bisphosphate FBP.  The same dynamical 
effects of calcium and
FBP saturations reported for the original BPLS model are observed here. We find out, however, a novel non-stationary effect which could be, in principle,  physiologically  interesting: some response times of the model tend to increase considerably for high concentrations of calcium and/or FBP. In particular, the larger the concentrations
of ${\rm Ca}_{\rm c}$ and/or FBP, the larger the necessary time to attain homeostasis. 
\end{abstract}

\section{Introduction}

	The exchange of energy in cells is mostly mediated by ATP (adenosine triphosphate) molecules. Such molecules are produced in several processes in an eukaryotic cell, but the principal source of ATP is typically the oxidative phosphorylation process which takes place in   mitochondria.  	The mitochondrion is an organelle with two membranes, having, therefore, two distinct bulk regions: the intermembrane space and the mitochondrial matrix.  In the inner  membrane, there are plenty of protein transporters and ionic channels, some of which execute an active transport leading to a gradient of some ions and molecules \cite{guyton,lehninger}. 
The metabolic cascade that leads to the production of ATP in the mitochondrion starts in the cytoplasm. At first, glucose is transported from the extracellular medium into the cytoplasm by GLUT transporters. It is then converted in glucose-6-phosphate (G6P) by the enzyme hexokinase. G6P is then converted in pyruvate in a process called glycolysis, in which there is a net production of 
two  ATP molecules. The pyruvate produced is transported into the mitochondrion (to the mitochondrial matrix) and is metabolized in a series of oxidation-reduction reactions in the citric acid cycle leading to the production of the   nicotinamide adenine dinucleotide  NAD  and flavin adenine dinucleotide  FAD.  These  electron donor molecules are oxidized in the complexes I to IV present in the inner mitochondrial membrane. These reactions lead to the activation of a proton pump, creating a pH gradient between the inter membrane space and the matrix. The protons pumped into the intermembrane space return to the matrix through a transporter that uses their energy to catalyze the conversion of ADP (adenosine diphosphate) into ATP. The ATP produced in the mitochondria is then transported to the cytoplasm by the ATP/ADP exchanger \cite{guyton,lehninger}.  

The kinetic aspects of the processes 
involved in the ATP production in mitochondria are rather intricate.
This issue was addressed by Magnus and Keizer (MK), who introduced in the series of papers
  \cite{mk_1997,mk_1998_a,mk_1998_b}
 a theoretical kinetic model for ATP production in mitochondria   based on the known biophysical properties of the enzymes and transporters involved in the process.  In fact,
the MK model was built  by considering electrical activity and cytosolic calcium handling
 in  insulin-secreting pancreatic $\beta$-cells. The model consists basically in a set of equations describing the dynamics of the citric acid cycle, the proton pump, and the 
inner mitochondrial membrane transporters of  
 ATP and calcium.
 The MK model is effectively based on first  biophysical principles and provides a very detailed and accurate description of the processes considered  to be important for mitochondrial oxidative phosphorylation. However, it is also a rather complex model with
 cumbersome equations, preventing a systematic mathematical study of its main dynamical and physiological  properties.
 
A simplification of the MK model aiming to retain its main dynamical properties was introduced recently 
by Bertram, Pedersen, Luciani, and Sherman (BPLS) in   \cite{bertram}.
The BPLS model incorporates some refinements 
 introduced by Cortassa {\em et al.} in \cite{cortassa}
for the description of the ATP production in cardiac cells. In fact, BPLS model can be considered as an approximation of the Cortassa {\em et al.}'s model instead
of the original MK one. As we will see, this was probably the origin of some inaccuracies in the 
BPLS equations. 
 As in the original MK model, the
mitochondrial ATP production in the  BPLS model is governed by four dynamical variables,
namely the potential drop in the inner membrane
$\Delta V$  and the mitochondrial concentrations of:  reduced nicotinamide adenine dinucleotide  NADH,  adenosine diphosphate ADP, 
and   calcium Ca$_{\rm m}$.
 The mitochondrial concentrations of  pyridine and adenine nucleotides are
assumed to be conserved
\begin{equation}
\label{NADtot}
{\rm NAD}_{\rm m} + {\rm NADH}_{\rm m} = {\rm NAD}_{\rm tot},
\end{equation}
\begin{equation}
\label{Atot}
{\rm ADP}_{\rm m} + {\rm ATP}_{\rm m} = {\rm A}_{\rm tot},
\end{equation}
where ${\rm NAD}_{\rm tot} $ and ${\rm A}_{\rm tot}  $ stand for the total mitochondrial concentration of the 
respective nucleotides. 
The balance of the pertinent fluxes and reactions yields to the following dynamical equations for the mitochondrial variables
\begin{eqnarray}
\label{e1}
\frac{d}{dt} {\rm NADH}&=&  J_{\rm PDH} - J_0  ,\\
\label{e2}
\frac{d}{dt} {\rm ADP}&=&  J_{\rm ANT} -J_{F1F0} ,\\
\label{e3}
\frac{d}{dt}Ca_{\rm m}&=& f_{\rm m}\left( J_{\rm uni} - J_{\rm NaCa} \right),\\
\label{e4}
 \frac{d}{dt} \Delta V&=& C_{\rm m}^{-1}\left(J_{H} 
- J_{\rm ANT} - J_{\rm NaCa} - 2J_{\rm uni}\right) ,
\end{eqnarray} 
where
\begin{equation}
J_{H} = J_{H,{\rm res}} -J_{H,{\rm ATP}} 
  - J_{H,{\rm leak}} .
\end{equation}
The derivation and meaning of the fluxes presented in the right-handed sides of Eq. (\ref{e1})-(\ref{e4}) are rather involved. The main details and the pertinent references can be found, for instance, in the BPLS paper \cite{bertram}. We have checked carefully the derivation of each of these fluxes and we have found out some inaccuracies in the BPLS expressions for the
adenine nucleotide translocator rate 
$J_{\rm ANT}$ and for the calcium uniporter rate 
$J_{\rm uni}$. As we will see, some of these problems probably have originated in the transcription
of the original MK equations to the  Cortassa {\em et al.}'s model.

In the present paper, we propose some enhanced approximations in the BPLS framework for the
fluxes  $J_{\rm ANT}$ and  $J_{\rm uni}$ 
 and analyze some of the dynamical properties of Eqs. (\ref{e1})-(\ref{e4}). We show, in particular, that for  
physiologically acceptable  ranges of mitochondrial respiration inputs, namely the cytosolic concentration of calcium ${\rm Ca}_{\rm c}$ and the substrate fructose 1,6-bisphosphate 
FBP, the BPLS equations have a unique physiologically acceptable attractor fixed point,
as one would indeed expect for any model compatible with homeostasis.
Exhaustive numerical explorations indicate that the BPLS model is indeed globally stable, reinforcing its relevance to physiological quantitative studies, despite its simplicity when compared to the MK original model. We 
 determine, in the stationary regime, the dependence on constant 
respiration inputs  
  ${\rm Ca}_{\rm c}$ and FBP of the   four  mitochondrial variables considered in the model. As in the original
 BPLS model, we observe here qualitatively distinct dynamical behavior 
for low and high concentrations of  ${\rm Ca}_{\rm c}$  and/or FBP. 
We detect, moreover, a  non-stationary effect which could be, in principle,  physiologically  interesting: the inertia of the system tends to increase considerably for high concentrations of cytosolic calcium and FBP, {\em i.e.},
  some response times of the model tend to increase considerably for high 
respiration inputs  
  ${\rm Ca}_{\rm c}$ and FBP. In particular, the larger the concentrations
of ${\rm Ca}_{\rm c}$ and/or FBP, the larger the necessary time to attain homeostasis.

\section{The Enhanced BPLS Model}

We will focus here in the problems we found for the BPLS 
 expressions for the 
adenine nucleotide translocator rate 
$J_{\rm ANT}$ and for calcium uniporter rate 
$J_{\rm uni}$, since  all the
 other quantities  appearing in (\ref{e1})-(\ref{e4}) were checked to be correct and
accurate for physiological ranges of variables and parameters. The MK expression for the 
former is (see Eq. (16) and Table 4 of \cite{mk_1997})
\begin{equation}
\label{JANT}
J_{\rm ANT}=V_{\rm max,ANT}\frac{1-\frac{\alpha_{\rm c}}{\alpha_{\rm m}}\frac{\rm ATP_{\rm c}}{\rm ADP_{c}}\frac{\rm ADP_{m}}{\rm ATP_{m}}e^{-\frac{F\Delta V}{RT}}}{\left(1+\alpha_{\rm c}\frac{\rm ATP_{c}}{\rm ADP_{c}}e^{-f\frac{F\Delta V}{RT}}\right)\left(1+\alpha_{\rm m}^{-1}\frac{\rm ADP_{m}}{\rm ATP_{m}}\right)}.
\end{equation}
The precise meaning of all the quantities presented in this formula can be found in \cite{mk_1997,mk_1998_a,mk_1998_b}, and in the BPLS paper \cite{bertram} as well. (For the values of the parameters, see Table 1). On the other
hand, the expression for $J_{\rm ANT}$  presented in the Eq. (35) of the Cortassa {\em et al.} paper  \cite{cortassa} reads
\begin{equation}
\label{JANTerr}
J_{\rm ANT}=V_{\rm max,ANT}\frac{1-\frac{\alpha_{\rm c}}{\alpha_{\rm m}}\frac{\rm ATP_{\rm c}}{\rm ADP_{c}}\frac{\rm ADP_{m}}{\rm ATP_{m}}}{\left(1+\alpha_{\rm c}\frac{\rm ATP_{c}}{\rm ADP_{c}}e^{-f\frac{F\Delta V}{RT}}\right)\left(1+\alpha_{\rm m}^{-1}\frac{\rm ADP_{m}}{\rm ATP_{m}}\right)}.
\end{equation}
By comparing with (\ref{JANT}), we see clearly that 
it  
lacks the exponential in the numerator. Furthermore, the incorrect expression
(\ref{JANTerr}) is transcribed in the BPLS Eq. (35) as
\begin{equation}
\label{JANTerr1}
J_{\rm ANT}=V_{\rm max,ANT}\frac{\frac{\rm ATP_{m}}{\rm ADP_{m}}-\frac{\alpha_{\rm c}}{\alpha_{\rm m}}\frac{\rm ATP_{\rm c}}{\rm ADP_{c}}}{\left(1+\alpha_{\rm c}\frac{\rm ATP_{c}}{\rm ADP_{c}}\right)\left(\frac{\rm ATP_{m}}{\rm ADP_{m}}+\alpha_{\rm m}^{-1} \right)e^{-f\frac{F\Delta V}{RT}}},
\end{equation} 
{\em i.e.},
with another  mistake in the denominator. 
The BPLS expression for $J_{\rm ANT}$, obtained from (\ref{JANTerr1}) after some simplifications, is
\begin{equation}
\label{JANT0}
J_{\rm ANT}=p_{19}\left(\frac{\frac{\rm ATP_{m}}{\rm ADP_{m}}}{\frac{\rm ATP_{m}}{\rm ADP_{m}}+p_{20}}\right)e^{f\frac{F\Delta V}{RT}},
\end{equation}
where $p_{19}$ and $p_{20}$ are some (fitted) numerical parameters.
The (reasonable) physiological hypothesis
used to derive (\ref{JANT0}) in the BPLS model is the assumption that, due to the ion transporters action, the rates of ATP to ADP in the mitochondrial matrix and in the cytoplasm are approximately the same,
\begin{equation}
\label{approx}
\frac{\rm ATP_{c}}{\rm ADP_{c}} \approx \frac{\rm ATP_{m}}{\rm ADP_{m}}.
\end{equation}
Note that this assumption implies from (\ref{JANT}) that $J_{\rm ANT}\approx 0$ for 
 ${\rm ATP_m} \to {\rm A_{tot}}$ (and, hence, ${\rm ADP_m} \to 0$ according to (\ref{Atot})), which is incompatible with the BPLS expression (\ref{JANT0}). 
 Another qualitatively different behavior arises for larges values of  
  $\Delta V$: equation (\ref{JANT}) implies that $J_{\rm ANT}$ tends to an asymptote, whereas (\ref{JANTerr1}) suggests an exponential growth. The expression (\ref{JANTerr1})
  is clearly not accurate as an approximation of (\ref{JANT}).

\begin{figure}[t!]
    \includegraphics[width=\linewidth]{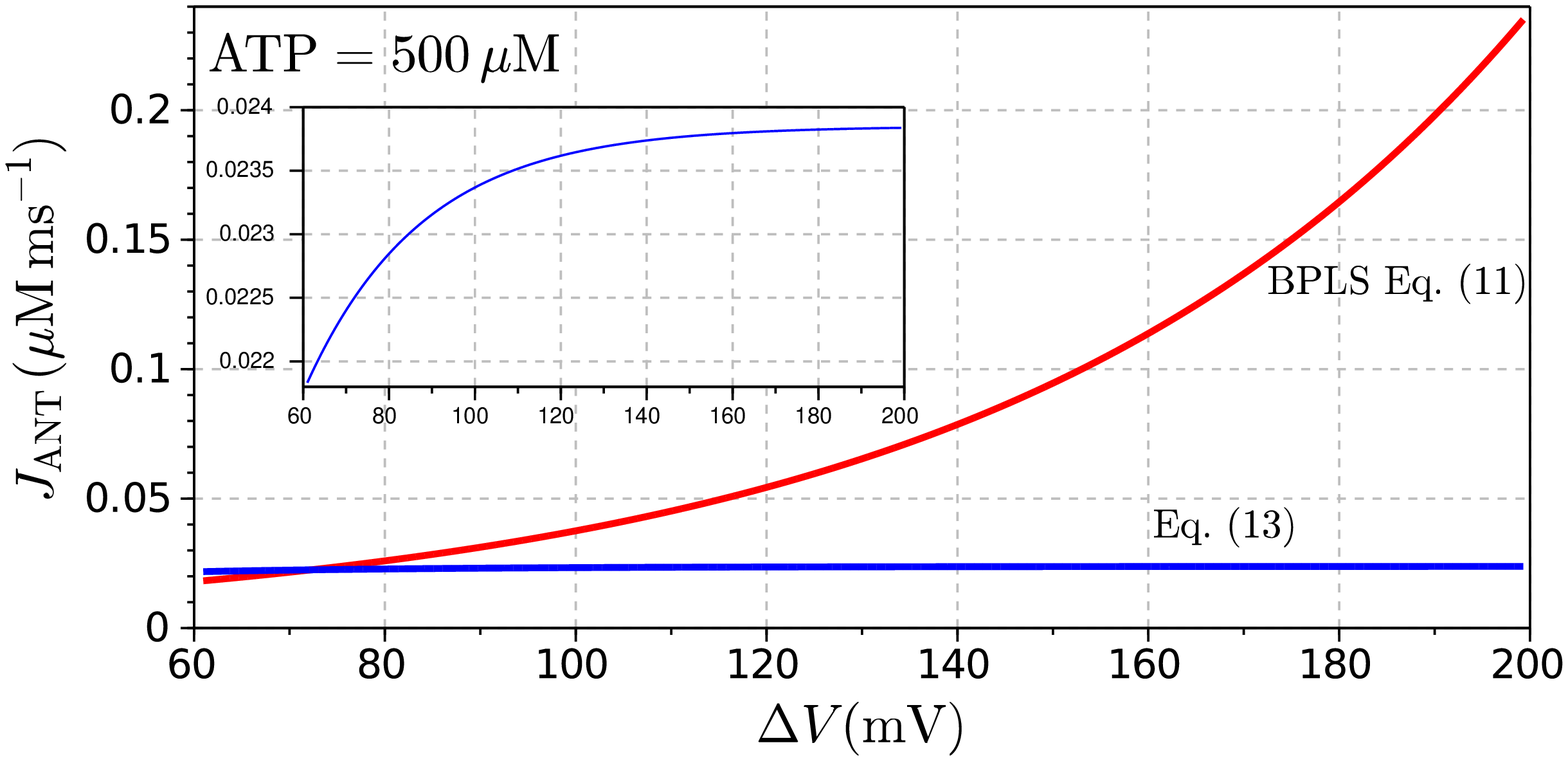}\\
  
 \includegraphics[width=\linewidth]{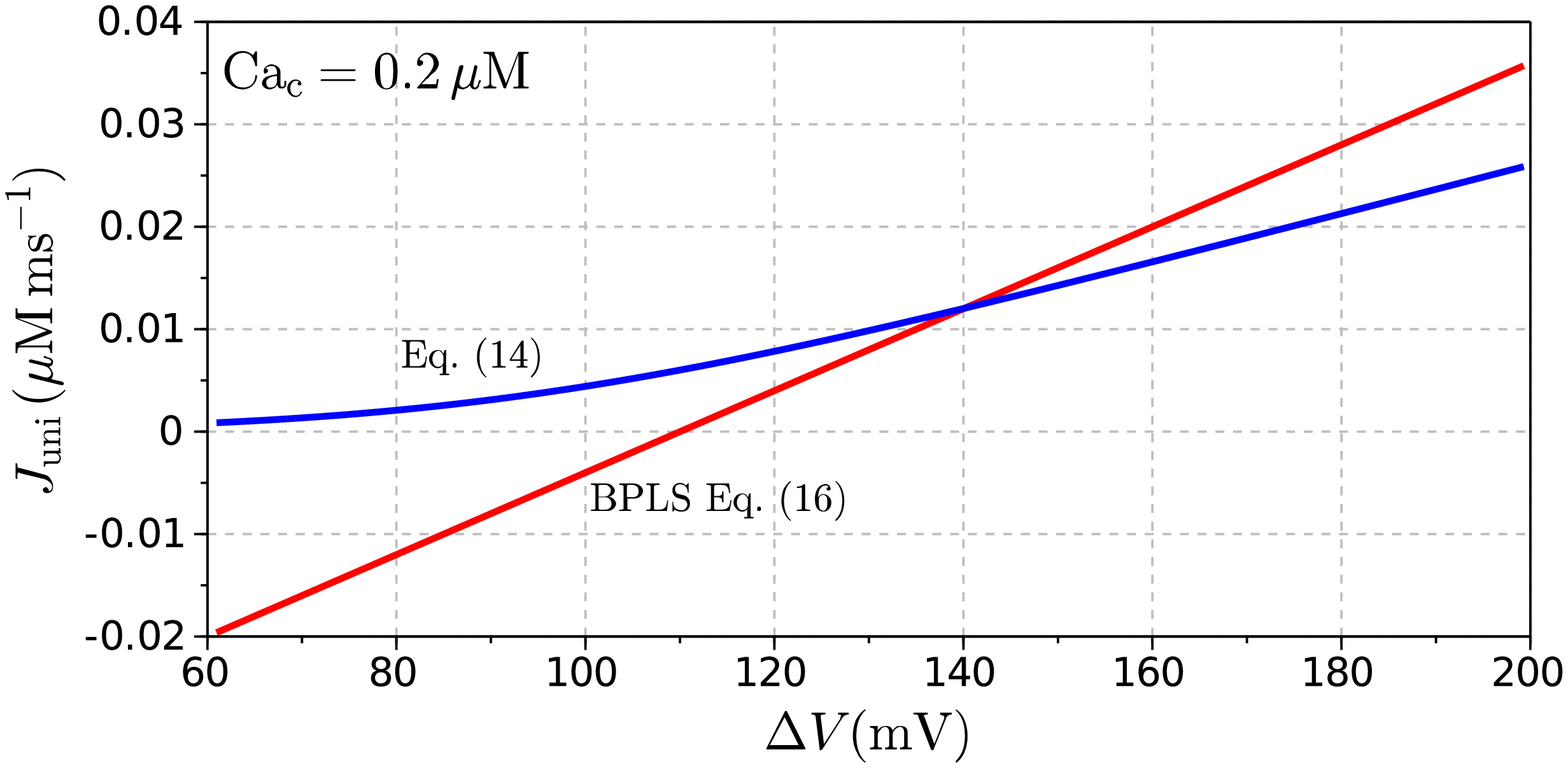}
\caption{Comparison between the original BPLS expressions and our proposals based on the original MK model.
Above: The BPLS  
adenine nucleotide translocator rate 
$J_{\rm ANT}$  (\ref{JANT0}) and our proposal (\ref{JANT1}). We notice that the
variation of 
 (\ref{JANT1})  over physiological ranges (the inserted graphics) is considerable
smaller than that one of (\ref{JANT0}). Furthermore, 
even the concavities of the curves are different. The dependency of  (\ref{JANT0}) on $\Delta V$ is exponential, whereas (\ref{JANT1}) tends to an asymptote for large
values of $\Delta V$.  In  accordance to the Table 1,
these curves were calculated by assuming ${\rm ATP} = 500\, \mu$M and ${\rm A}_{\rm tot} = 15\,$mM.
Below: The BPLS   calcium uniporter rate 
$J_{\rm uni}$ (\ref{Juni1}) and the  original MK expression (\ref{Juni}), both calculated for ${\rm Ca}_{\rm c} = 0.2\, \mu$M. Eq. (16) is a straight line which implies  non-positive rates for physiological values of $\Delta V$.
 }
\label{fig1}
\end{figure}

With the  assumption (\ref{approx}) and taking into account the conservation of mitochondrial pyridine nucleotides (\ref{Atot}), the original 
MK expression (\ref{JANT}) for the
adenine nucleotide translocator rate reads
\begin{equation}
\label{JANT1}
J_{\rm ANT}=V_{\rm max,ANT}\left(\frac{\rm  ATP_{m}}{{\rm A}_{\rm tot} - (1-\alpha_{\rm m}){\rm  ATP_{m}}} \right)
\frac{\alpha_{\rm m}-\alpha_{c}e^{-\frac{F\Delta V}{RT}}}{1+\alpha_{\rm c}\frac{\rm ATP_{m}}{{\rm A}_{\rm tot} -  {\rm  ATP_{m}}}e^{-f\frac{F\Delta V}{RT}}}.
\end{equation}
This is our first proposed approximation, 
which captures 
  all the essential properties of (\ref{JANT}) and is still simple enough to be
mathematically   
   manipulated. Notice that for the typical range of physiological parameters, neglecting   the exponential in the numerator of (\ref{JANT1}) would imply a relative error inferior to 5\%. We will not, however, adopt this further approximation in this work.
Figure (\ref{fig1})  
illustrate the discrepancies between the expressions (\ref{JANT0}) and (\ref{JANT1}) for
typical physiological values of the parameters and variables.
A closer inspection of the graphics (9) of \cite{bertram} reveals that they have
probably  compared their approximated expression 
(\ref{JANT0})
with  Eq. (\ref{JANTerr1}), which was itself transcribed incorrectly from  Cortassa {\em et al.}'s Eq; (\ref{JANTerr}).

With respect to the calcium uniporter rate 
$J_{\rm uni}$, the original MK expression reads (see Eq. (19) in \cite{mk_1997})
\begin{equation}
\label{Juni}
J_{\rm uni}=V_{\rm max,uni}
\frac{\frac{2F}{RT}(\Delta V-\Delta V_{0})}{1-e^{-\frac{2F}{RT}(\Delta V-\Delta V_{0})}} 
\left(\frac{\frac{\rm Ca_{c}}{K_{\rm trans}}\left(1+\frac{\rm Ca_{c}}{K_{\rm trans}}\right)^{3}}{\left(1+\frac{\rm Ca_{c}}{K_{\rm trans}}\right)^{4}+\frac{L}{(1+{\rm Ca_{c}}/K_{\rm act})^{n_{a}}}}\right).
\end{equation}
In the BPLS derivation of the approximation for $J_{\rm uni}$, it is used Eq. (38) of Cortassa {\em et al.} \cite{cortassa}, which reads
\begin{equation}
\label{Junierr}
J_{\rm uni}=V_{\rm max,uni} 
\frac{\frac{\rm Ca_{c}}{K_{\rm trans}}\left(1+\frac{\rm Ca_{c}}{K_{\rm trans}}\right)^{3}\frac{2F}{RT}(\Delta V-\Delta V_{0})}{\left(1+\frac{\rm Ca_{c}}{K_{\rm trans}}\right)^{4}+\frac{L}{(1+{\rm Ca_{c}}/K_{\rm act})^{n_{a}}}
\left(
1-e^{-\frac{2F}{RT}(\Delta V-\Delta V_{0})}\right)},
\end{equation}
 where one can see that there is a mistake in the denominator. The BPLS proposed expression for the calcium uniporter rate, obtained as a simplification of (\ref{Junierr}), is
\begin{equation}
\label{Juni1}
J_{\rm uni}=(p_{21}\Delta V-p_{22}){\rm Ca_{c}}^{2},
\end{equation}
where $p_{21} \approx 0.01\, \mu{\rm M}^{-1}{\rm ms}^{-1}{\rm mV}^{-1}$ and 
$p_{22} \approx 1.1 \, \mu{\rm M}^{-1}{\rm ms}^{-1}$ are also fitted numerical parameters.
We found 
this equation to be  inaccurate for the typical physiological range of parameters as well
(see Figure (\ref{fig1})). Notice, in particular, that it implies in   non-positive flux rates for $\Delta V \le p_{22}/p_{21} \approx 110 \,$mV.
 We propose to keep in the approximated model the complete original MK equation (\ref{Juni}). Its dependence on
$\Delta V$ is already in a rather simple form, and the complications for $\rm Ca_{c}$ are harmless for the dynamical studies, as we will show. 

For the dynamical analysis, it is more conveniently to introduce 
the following dimensionless variables
\begin{eqnarray}
\label{def1}
x &=& \frac{{\rm NADH}_{\rm m}}{{\rm NAD}_{\rm tot}}, \\
\label{def2}
y &=& \frac{{\rm ATP}_{\rm m}}{{\rm A}_{\rm tot}}, \\
\label{def3}
z &=& \frac{\rm Ca_m}{{\rm Ca}_0}, \\
\label{def4}
w &=& \frac{\Delta V}{\Delta V_0}, \\
\label{d1}
u &=& \frac{\rm Ca_c}{{\rm Ca}_0},\\
\label{d2}
v &=& {\frac{\rm FBP}{{\rm FBP}_0}}, 
\end{eqnarray}
Taking into account the new proposed expressions (\ref{JANT1}) and (\ref{Juni}),
the rates in the right-handed sides of (\ref{e1})-(\ref{e4}) will be given by
\begin{eqnarray}
\label{JPDH}
J_{\rm PDH} &=& r_1\sqrt{v} \frac{z}{a_1 + z}\left(a_2 + \frac{x}{1-x}\right)^{-1}  , \\
J_0 &=& r_2\frac{x}{a_3+x}\left(1+a_4e^{a_5w}\right)^{-1},\\
J_{\rm ANT} &=& r_3 \left(\frac{y}{1-a_6y}\right)\frac{a_7-a_8e^{-a_9w}}{1+a_8\frac{y}{1-y}e^{-a_{10}w}},\\
J_{F1F0} &=& r_4 \left[(a_{11}+y)\left(1 + a_{12}e^{-a_{13}w} \right) \right]^{-1},\\
J_{H,{\rm res}} &=& a_{14} J_0 , \\
J_{H,{\rm ATP}} &=& a_{15} J_{F1F0},\\
J_{H,{\rm leak}} &=& r_5 (w - a_{16}), \\
J_{\rm NaCa} &=& r_6\frac{z}{u}e^{a_{17}w}, \\ 
J_{\rm uni} &=&  r_7\frac{a_{18}(w-1)}{1-e^{-a_{18}(w-1)}}G(u),
\end{eqnarray}
where
\begin{equation}
G(u) =  \frac{u(1+a_{19}u)^{n_a}(1+a_{20}u)^3}{a_{21}+(1+a_{19}u)^{n_a}(1+a_{20}u)^4   }.
\end{equation}
All the values of the numerical parameters and constants are   presented in Table \ref{param}. 
\begin{table}
\begin{tabular}{lll}
\hline
\hline
${\rm NAD}_{\rm tot} = 10\times 10^3\, \mu{\rm M}\quad$ & ${\rm A}_{\rm tot} = 15\times 10^3\, \mu{\rm M}\quad$ & ${\rm Ca}_0 = 0.2\, \mu{\rm M}$\\
$\Delta V_0 = 91\, {\rm mV}$   & ${\rm FBP}_0= 1\, \mu{\rm M}$ & $C_m=1.8\, \mu{\rm M\, mV}^{-1} $\\
$V_{\rm max,ANT}= 5\, \mu{\rm M\, ms}^{-1} $ & 
$V_{\rm max,uni}= 10\, \mu{\rm M\, ms}^{-1} $ &  $\frac{F}{RT} = 0.037\, {\rm mV}^{-1} $ \\
$f=0.5$  &$K_{\rm trans} = 19\, \mu{\rm M}$  & $K_{\rm act} = 0.38\, \mu{\rm M}$ \\
$L=110$  & $f_m=0.01$& $n_a = 2.8$ \\ 
$\alpha_{\rm c} = 0.111$ & $\alpha_{\rm m} = 0.139$ &  \\
\hline
$r_1= 0.2\, \mu{\rm M\, ms}^{-1}$ & $r_2= 0.6\, \mu{\rm M\, ms}^{-1}$ & $r_3= 5\,\mu{\rm M\, ms}^{-1}$  \\ 
$r_4= 23.3\, \mu{\rm M\, ms}^{-1}$ &  $r_5= 0.182\, \mu{\rm M\, ms}^{-1}$ &  $r_6= 0.001\,\mu{\rm M\, ms}^{-1}$\\ 
 $r_7=  0.11\, \mu{\rm M\, ms}^{-1}$ &  & \\ 
\hline
$a_1 = 0.05 $ & $a_{2} = 1$ & $a_{3} = 0.01$ \\
$a_4 = 4.23\times 10^{-16}$ 	& $a_{5} = 18.2 $ & $a_{6} = 0.861$ \\ 
$a_7= 0.139 $ & $a_8 = 0.111$ & $a_9 =  3.37 $\\
$a_{10} = 1.68$ 	& $a_{11} = 0.67 $ & $a_{12} = 5.10\times 10^{9}$ \\
$a_{13} = 10.7 $ & $a_{14} =11.7 $ & $a_{15} = 3.43$ \\
$a_{16} = 0.16$ & $a_{17} = 1.46 $ & $a_{18} = 6.73 $ \\
$a_{19} = 0.52 $ & $a_{20}= 0.01$ & $a_{21} = 110$  \\
\hline
\hline
\end{tabular}
\caption{Numerical parameters and rates for the enhanced BPLS model, see equations (\ref{def1})-(\ref{ee4}). All the values were obtained from  \cite{mk_1997,mk_1998_a,mk_1998_b} and \cite{bertram}.\label{param}}
\end{table}
With the new dimensionless variables, the Eqs. (\ref{e1})-(\ref{e4}) can be cast in the form
\begin{eqnarray}
\label{ee1}
\dot{x} &=& \frac{1}{{\rm NAD}_{\rm tot}}\left( J_{\rm PDH} - J_0  \right) ,\\
\label{ee2}
\dot{y} &=&  \frac{1}{{\rm A}_{\rm tot}}\left(    J_{F1F0} -J_{\rm ANT}\right) ,\\
\label{ee3}
\dot{z}&=& \frac{f_{\rm m}}{{\rm Ca}_0}\left( J_{\rm uni} - J_{\rm NaCa} \right),\\
\label{ee4}
 \dot{w} &=& \frac{1}{C_{\rm m}\Delta V_0} \left(J_{H} 
- J_{\rm ANT} - J_{\rm NaCa} - 2J_{\rm uni}\right) ,
\end{eqnarray} 
where $f_{\rm m}$ and $C_{\rm m}$ stand, respectively, for
the fraction of free Ca ions and the mitochondrial capacitance, see Table 1. 
Equations (\ref{ee1})-(\ref{ee4}) form a non-autonomous systems of four first order differential equations. The external excitations $u(t)$ and $v(t)$ are related, respectively,
to the cytosolic concentration of calcium $\rm Ca_{\rm c}$ and the substrate fructose 1,6-bisphosphate 
FBP, see Eqs. (\ref{d1}) and (\ref{d2}). We can now start the dynamical analysis of the model.

\section{Dynamics of the model}

\begin{figure}[tb]
  \includegraphics[width=0.5\linewidth]{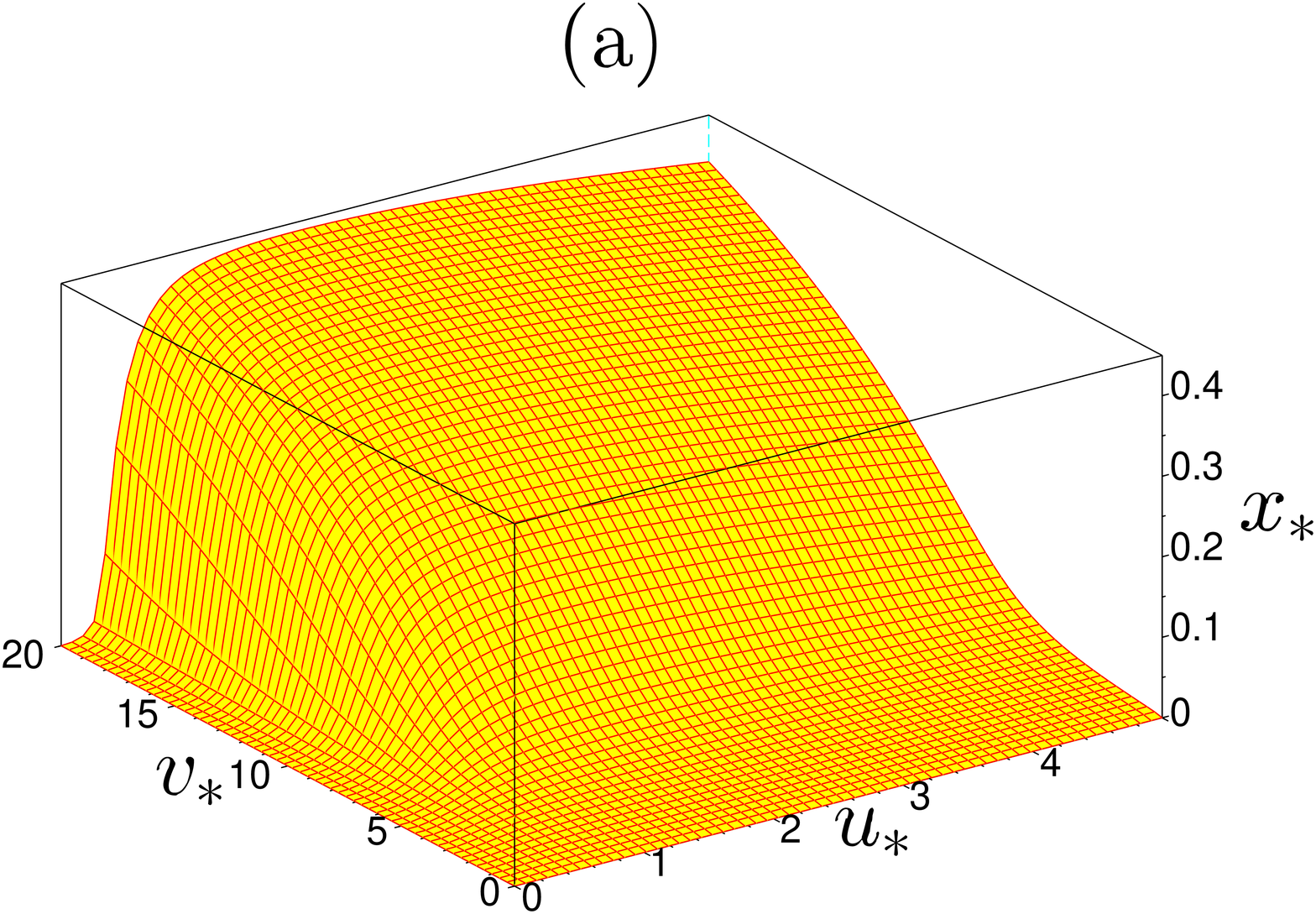}
  \includegraphics[width=0.5\linewidth]{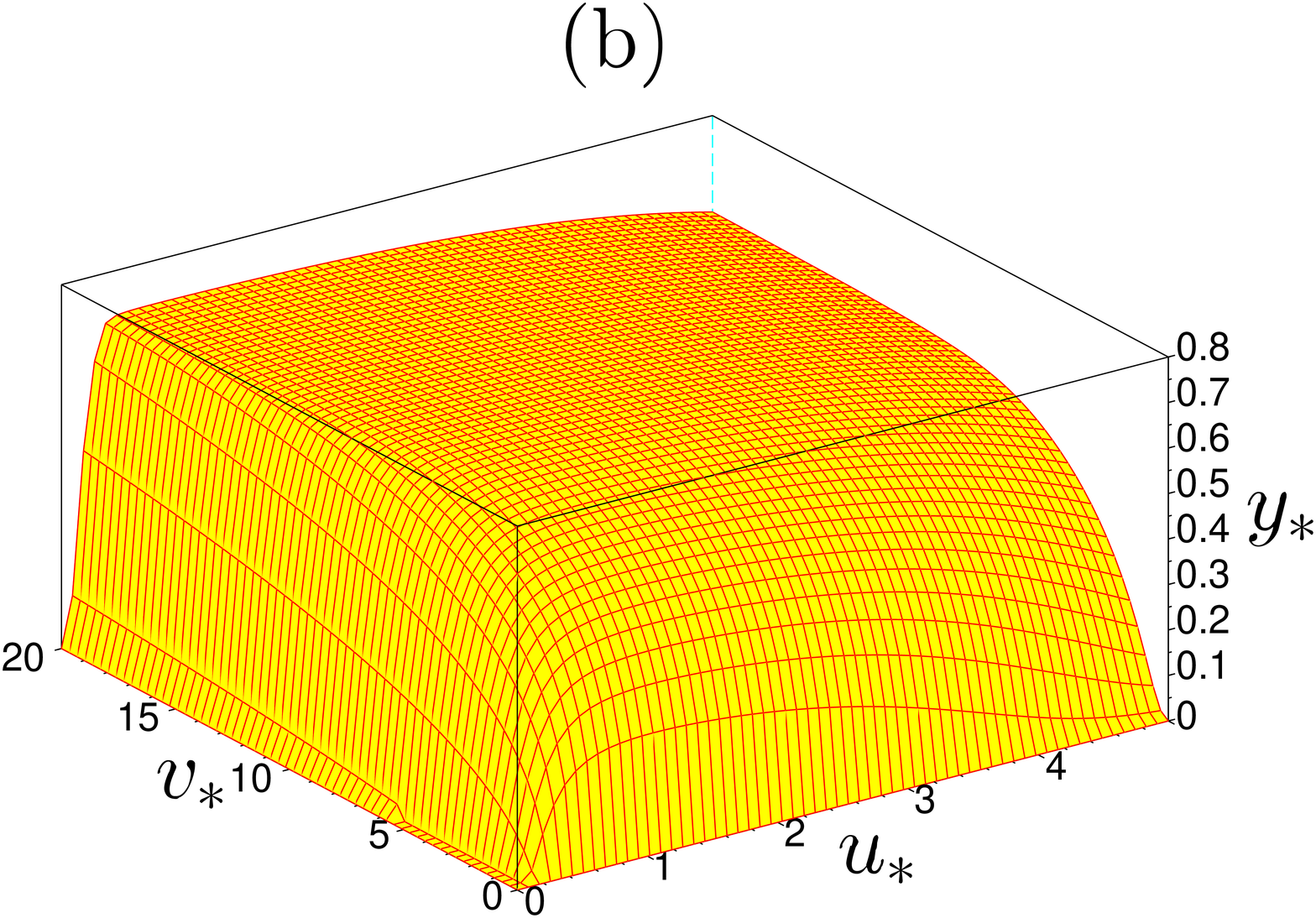}
  \includegraphics[width=0.5\linewidth]{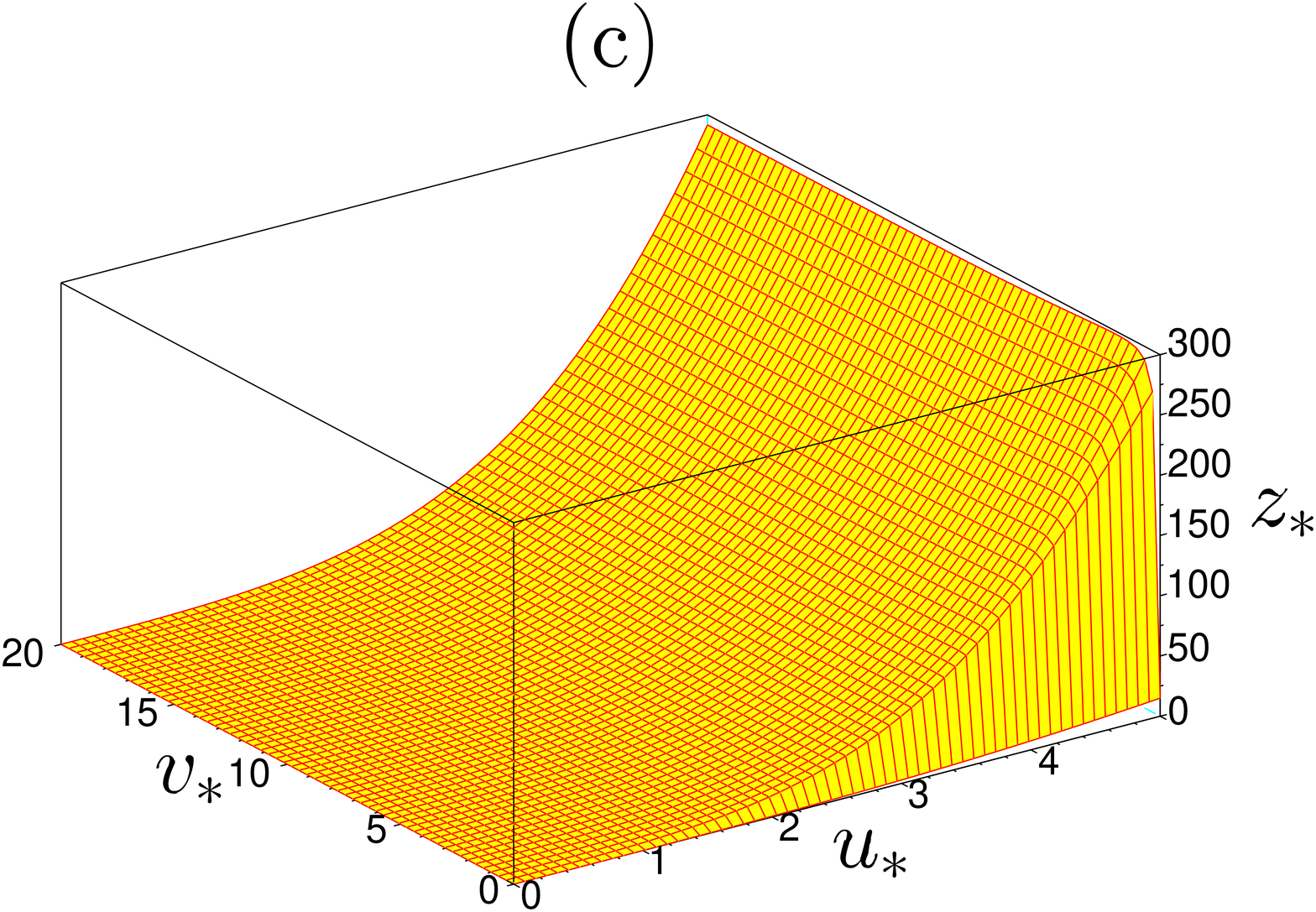}
  \includegraphics[width=0.5\linewidth]{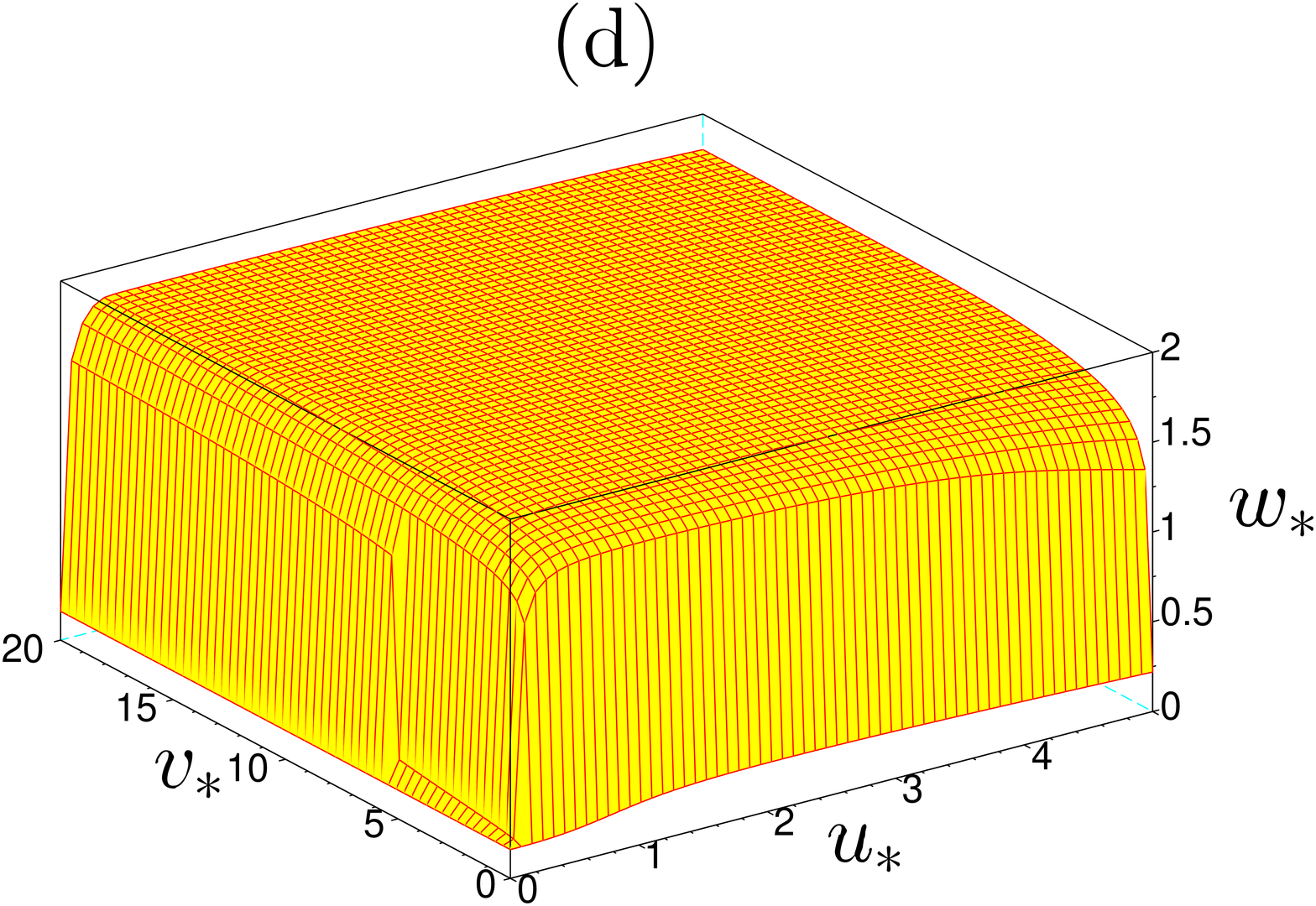}
\caption{The globally stable 
physiological fixed point $(x_*,y_*,z_*,w_*)$ of the system (\ref{ee1})-(\ref{ee4}) assuming constant inputs $(u_*,v_*)$.
 }
\label{fig2}
\end{figure}

Let us consider initially the 
 fixed points $(x_*,y_*,z_*,w_*)$ of the system (\ref{ee1})-(\ref{ee4}) assuming constant inputs $(u_*,v_*)$. By construction, the physiologically meaningful range for the variables $x$ and $y$ is   $[0,1]$, see (\ref{NADtot})-(\ref{Atot}) and (\ref{def1})-(\ref{def2}). For $z$ and $w$, we assume only that they are non negative. The typical 
 physiological range for the potential drop, however, is more restrictive, corresponding to $\Delta V \approx [90,225]\,$mV, which is 
 equivalent to  
 $w\approx [1,2.5]$. For the inputs $u$ and $v$, we consider the ranges $[0,10]$ and
 $[0,20]$, respectively, which corresponds to ${\rm Ca}_{\rm c}\approx [0,2]\,\mu$M  
 and ${\rm FBP}\approx [0,20]\, \mu$M. We perform an exhaustive numerical search \cite{scilab} for fixed points of
 (\ref{ee1})-(\ref{ee4}) by assuming $u\in [0,10]$ and $v\in[0,20]$ constants. For all tested values of $u$ and $v$, only one physiological ($x,y\in[0,1]$ both $z,w>0$) fixed point was found, which is always stable. Moreover, the fixed point is globally stable for physiological ranges of variables, meaning that any solution of (\ref{ee1})-(\ref{ee4}) with reasonable initial conditions will tend asymptotically to the fixed point,
 {\em i.e.}, the system indeed exhibits an asymptotic behavior 
compatible with homeostasis. Starting at a random point in the phase space, 
 the variables $w$ and $x$ have typically the quickest convergence to the fixed point, where $y$ and $z$ are the slowest ones.
The values of $(x_*,y_*,z_*,w_*)$  as function of the
 constant inputs $(u_*,v_*)$ are depicted in Fig. (\ref{fig2}), from where
one can already observe some physiologically consistent  dynamical properties which we describe in detail below.

The first observations is that the production of ATP  and the concentration of NADH vanishes in the absence of cytosolic  calcium ${\rm Ca}_{\rm c}$ and/or the substrate fructose 1,6-bisphosphate FBP, {\em i.e.},
$x_*$ and $y_* \to 0$ for $u_*$ or $v_*\to 0$. 
Notice that, from the condition
$J_{\rm NaCa} = J_{\rm uni}$ defining the fixed point $\dot z_*=0$ (see Eq. (\ref{ee3})), we have $z_*=0$ for $u_*=0$. On the other hand, $J_{\rm PDH}$ vanishes for $z_*=0$ (and for
$v_*=0$ as well), which implies via the condition $\dot x_*=0$ that $J_0=0$ and, consequently, $x_*=0$. The condition for $y_*$ and $w_*$ are more involved. The former
 vanishes for vanishing $u_*$ or $v_*$, while the latter will be given by 
$w*\approx a_{16}=0.16$ for $u_*=0$.
Also, we see that for reasonable values of $u_*$ and $v_*$ the value of the potential drop $\Delta V$ ($w_*$) is almost constant and close to
$150\,$mV ($w_*=1.65$). This stability is probably the reason why the original BPLS model is robust, despite the inaccuracies for the expression of $J_{\rm ANT}$ and $J_{\rm uni}$ we are correcting in this paper. We will return to this point in the last section.
Still from the condition $J_{\rm NaCa} = J_{\rm uni}$, we see that 
$z_*   \propto  u_*G(u_*)$, since $w_*$ is almost constant for physiological reasonable values of $u_*$ and $v_*$ (see Fig. (2c)).

\begin{figure}[t!]
  \includegraphics[width=0.5\linewidth]{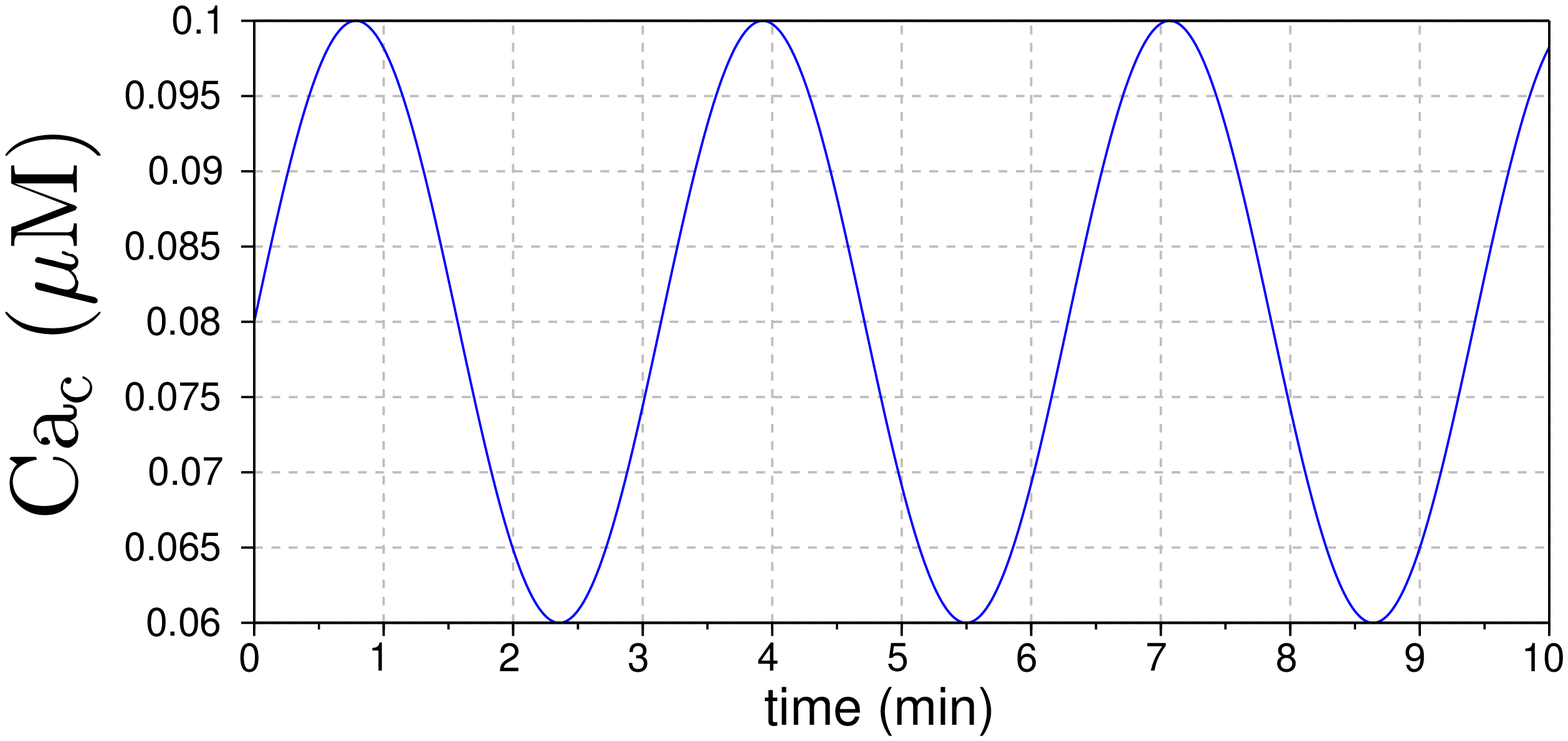}
  \includegraphics[width=0.5\linewidth]{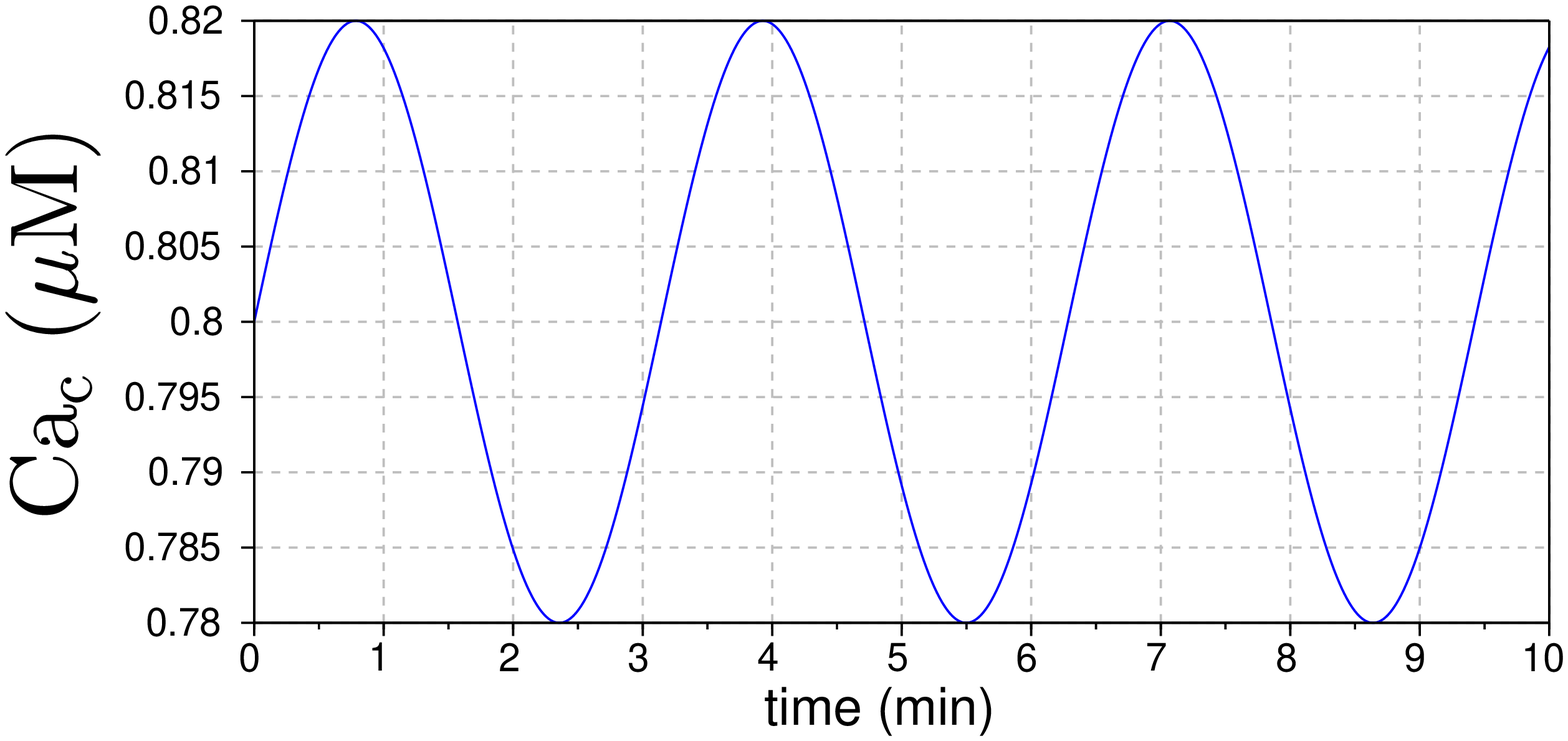}
  \includegraphics[width=0.5\linewidth]{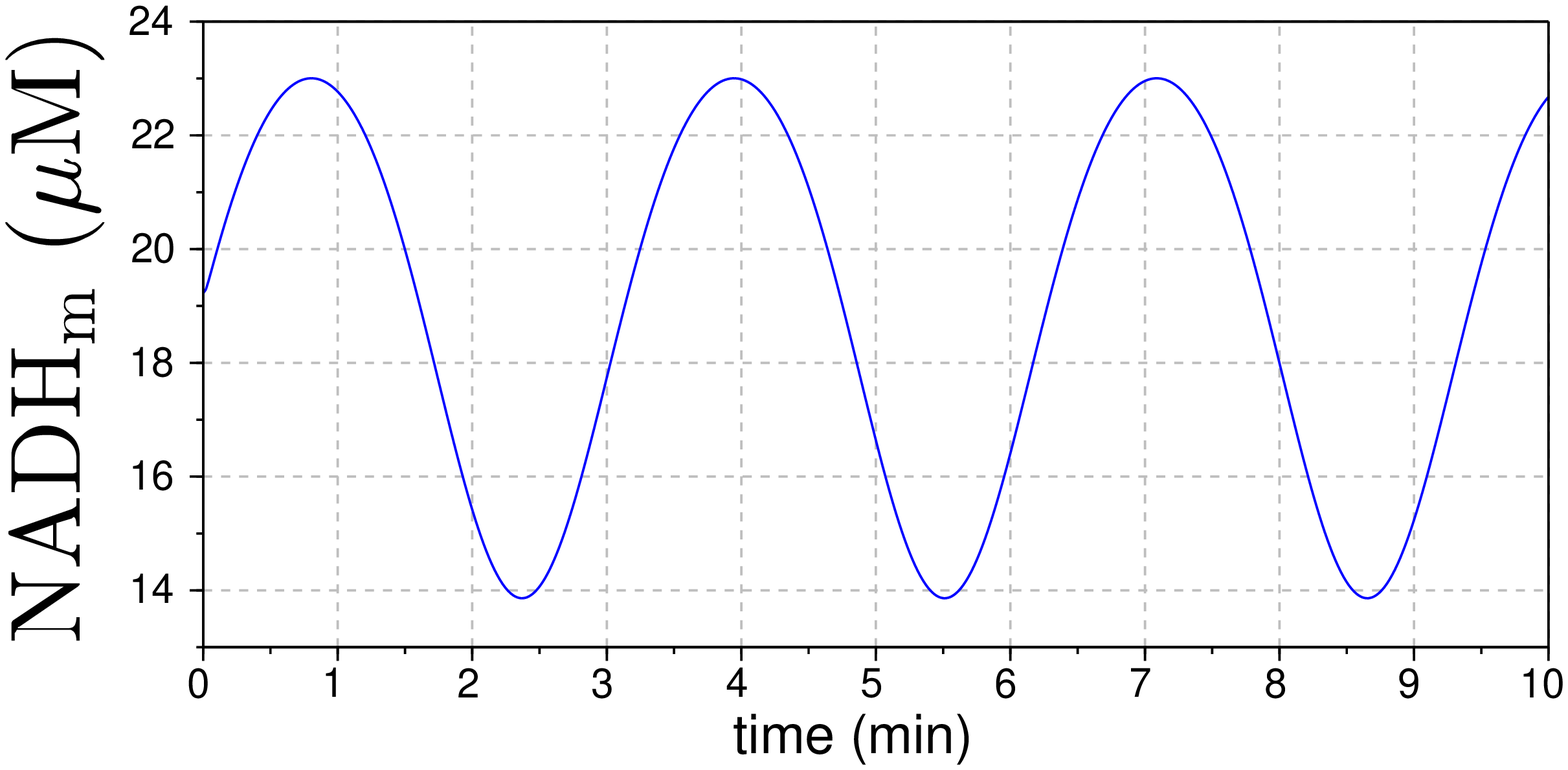}
  \includegraphics[width=0.5\linewidth]{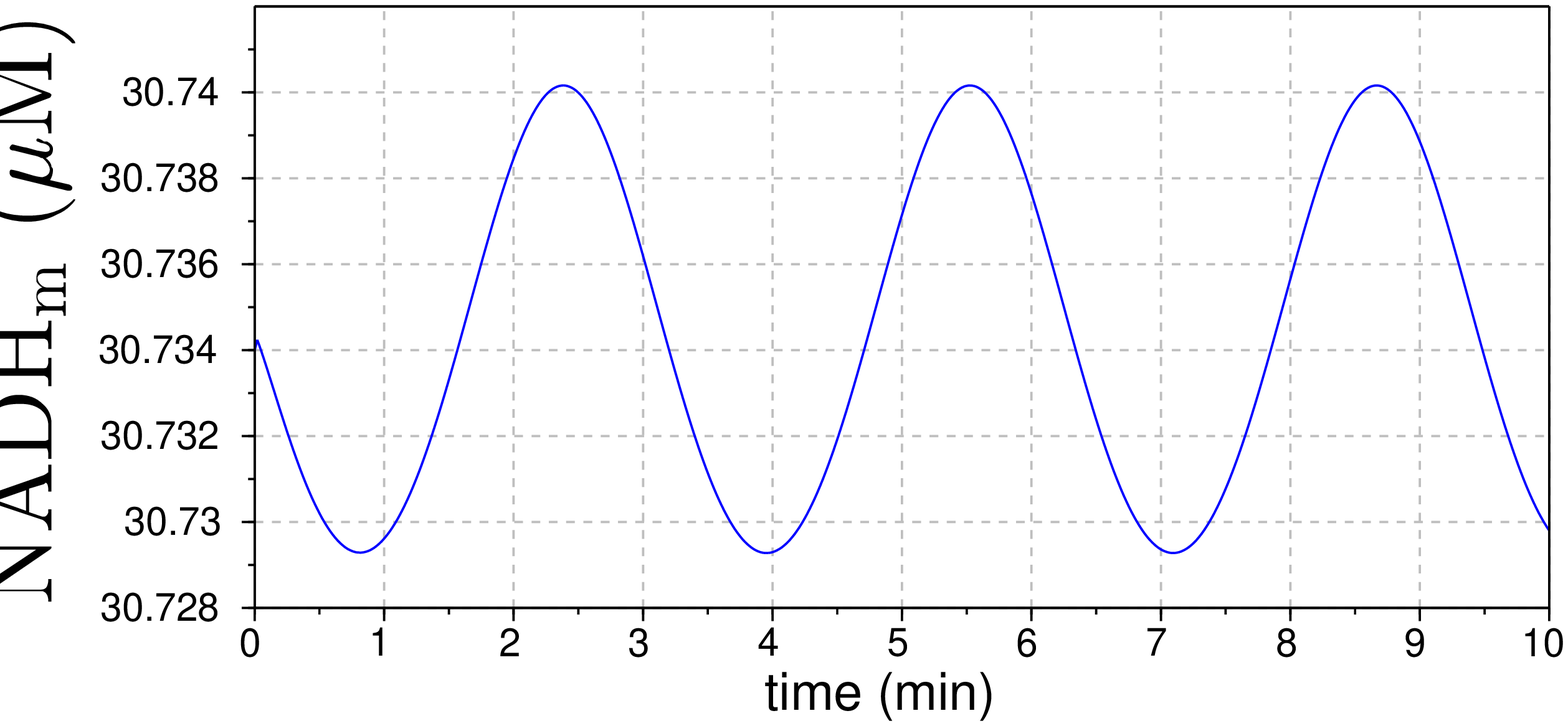}
  \includegraphics[width=0.5\linewidth]{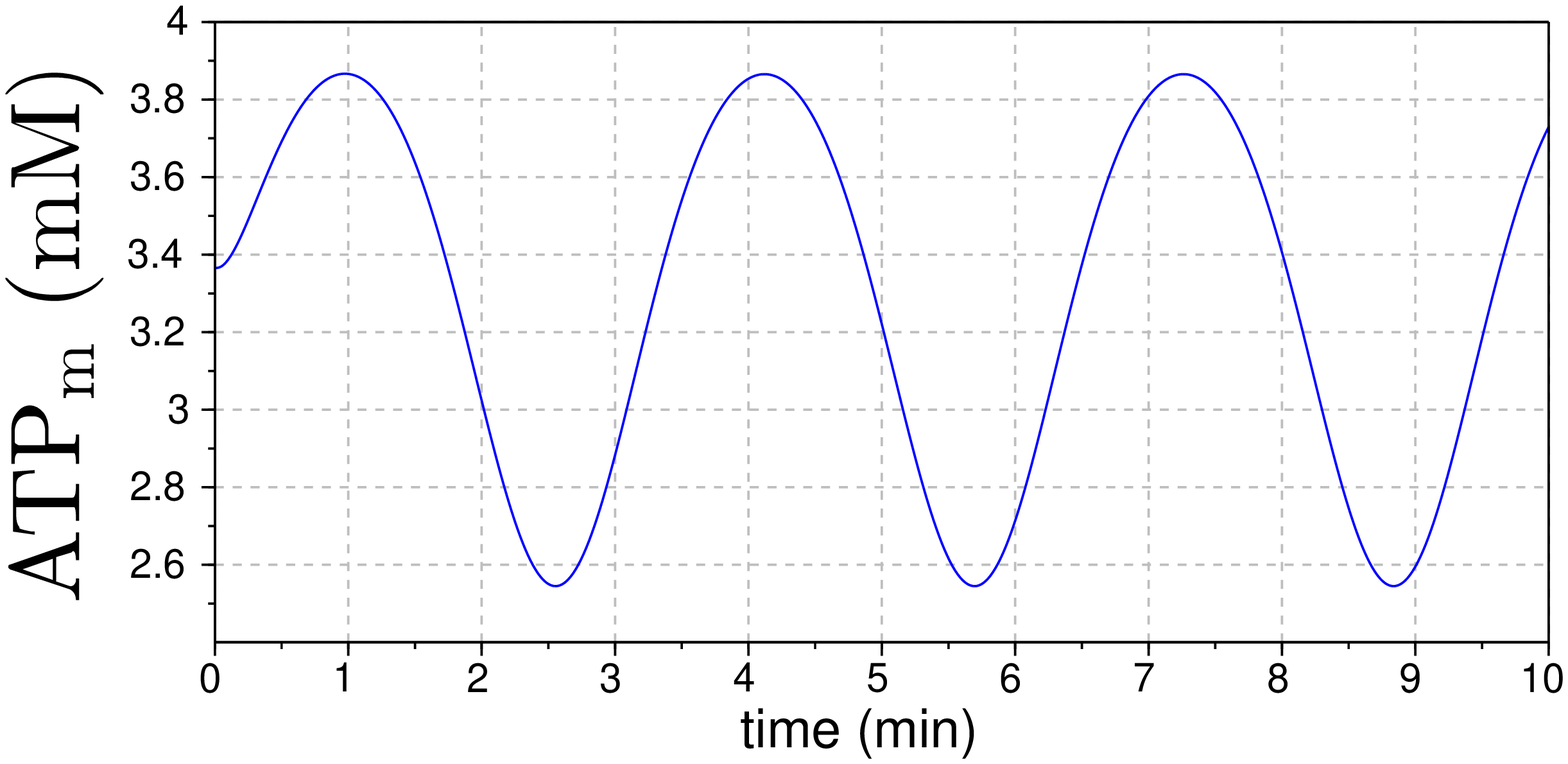}
  \includegraphics[width=0.5\linewidth]{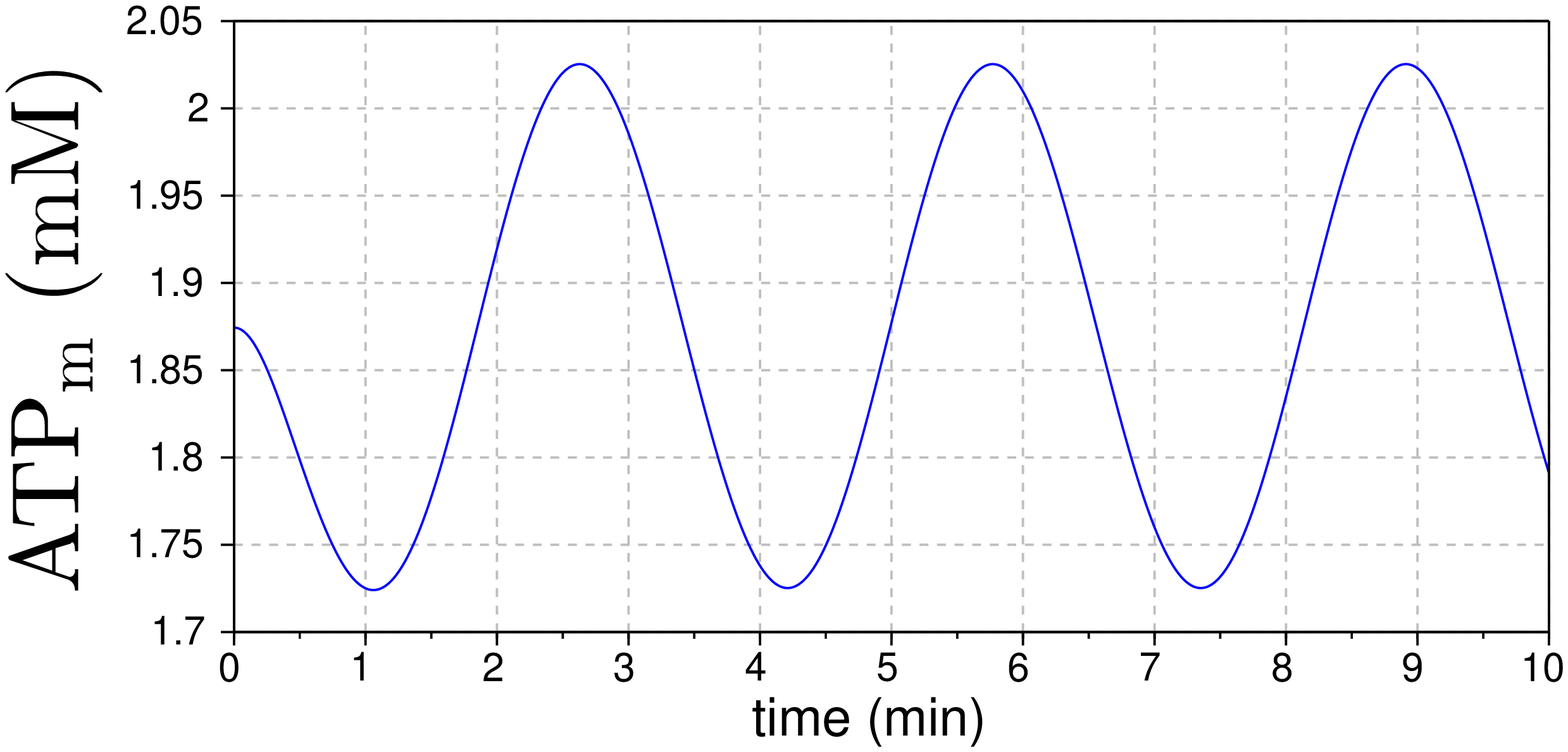}
  \includegraphics[width=0.5\linewidth]{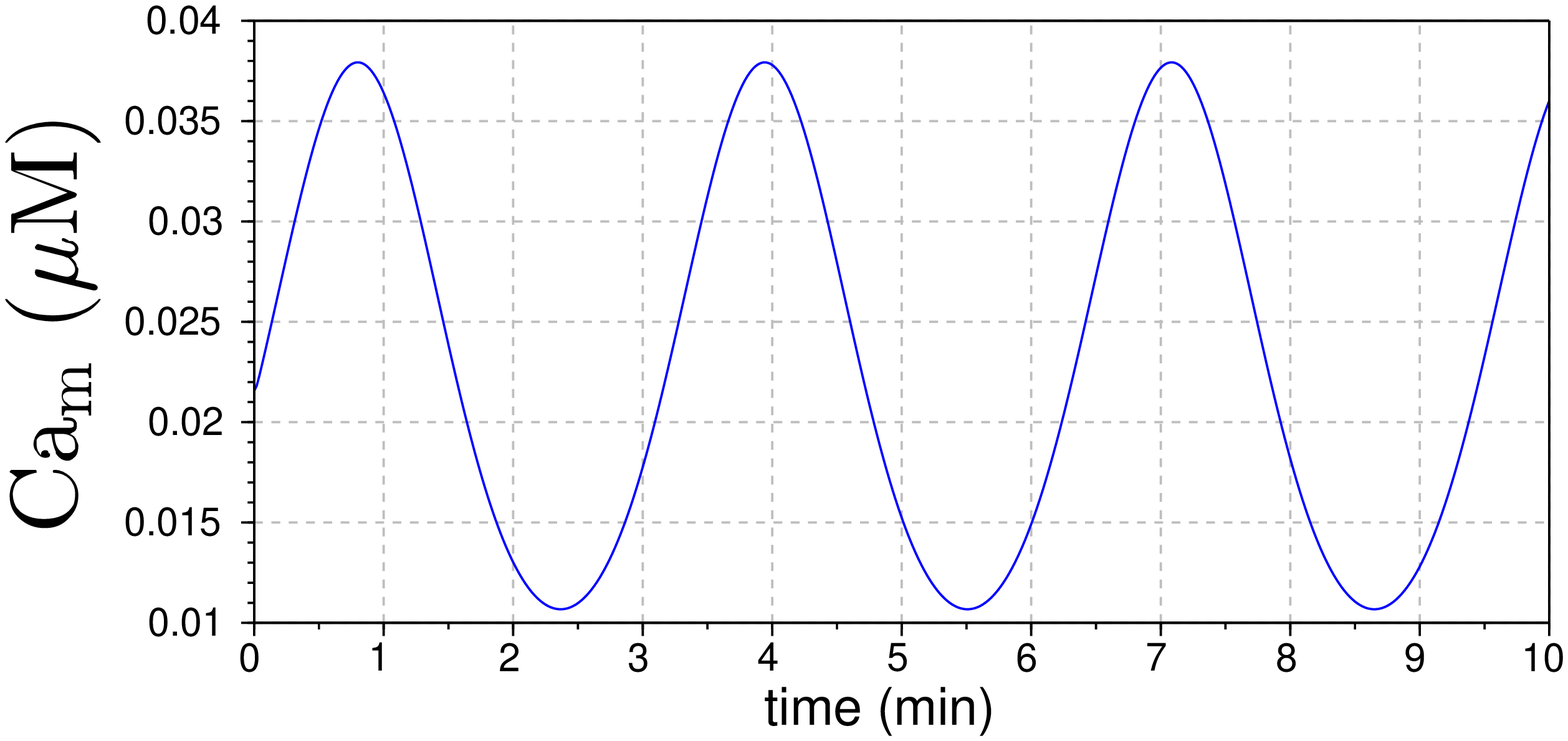}
  \includegraphics[width=0.5\linewidth]{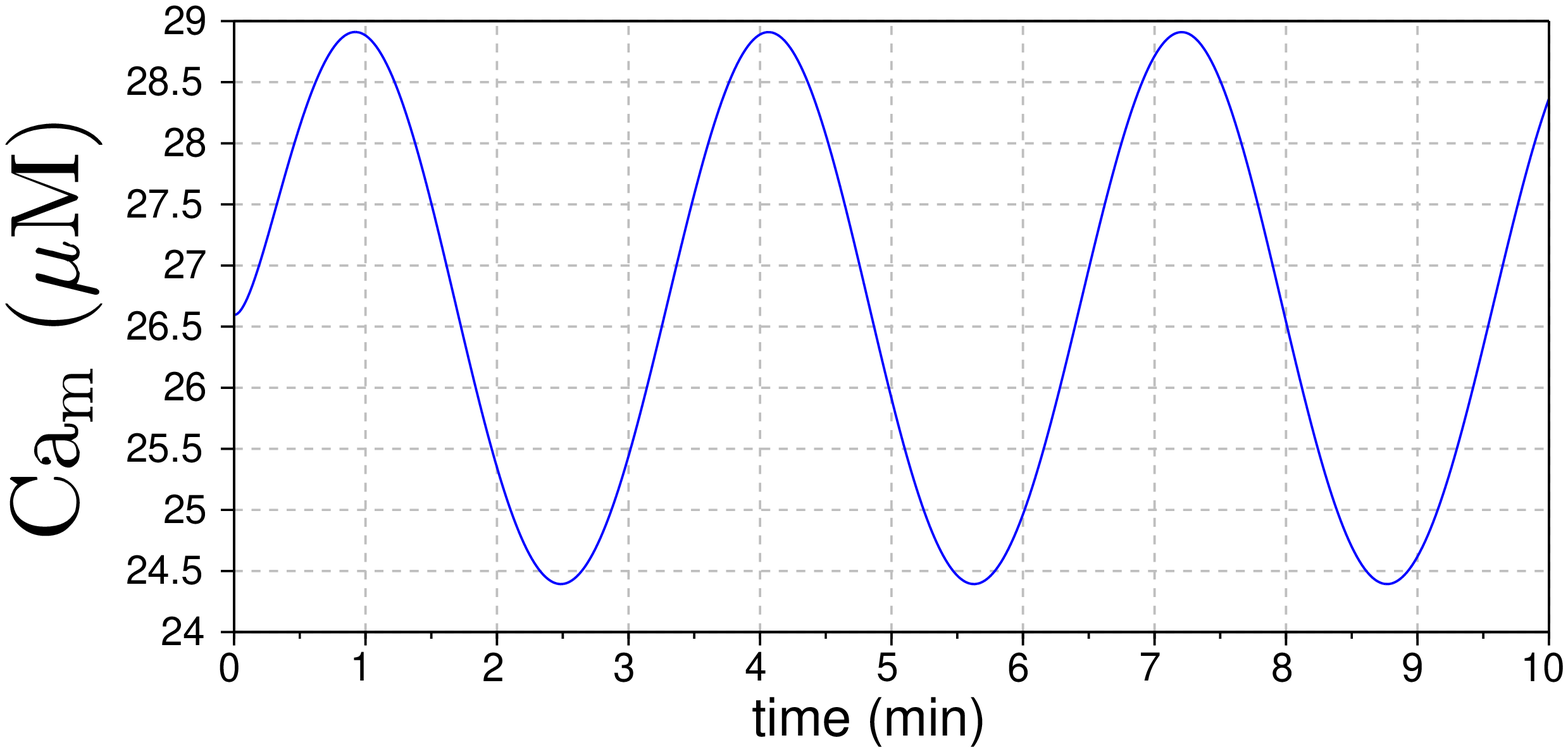}
  \includegraphics[width=0.5\linewidth]{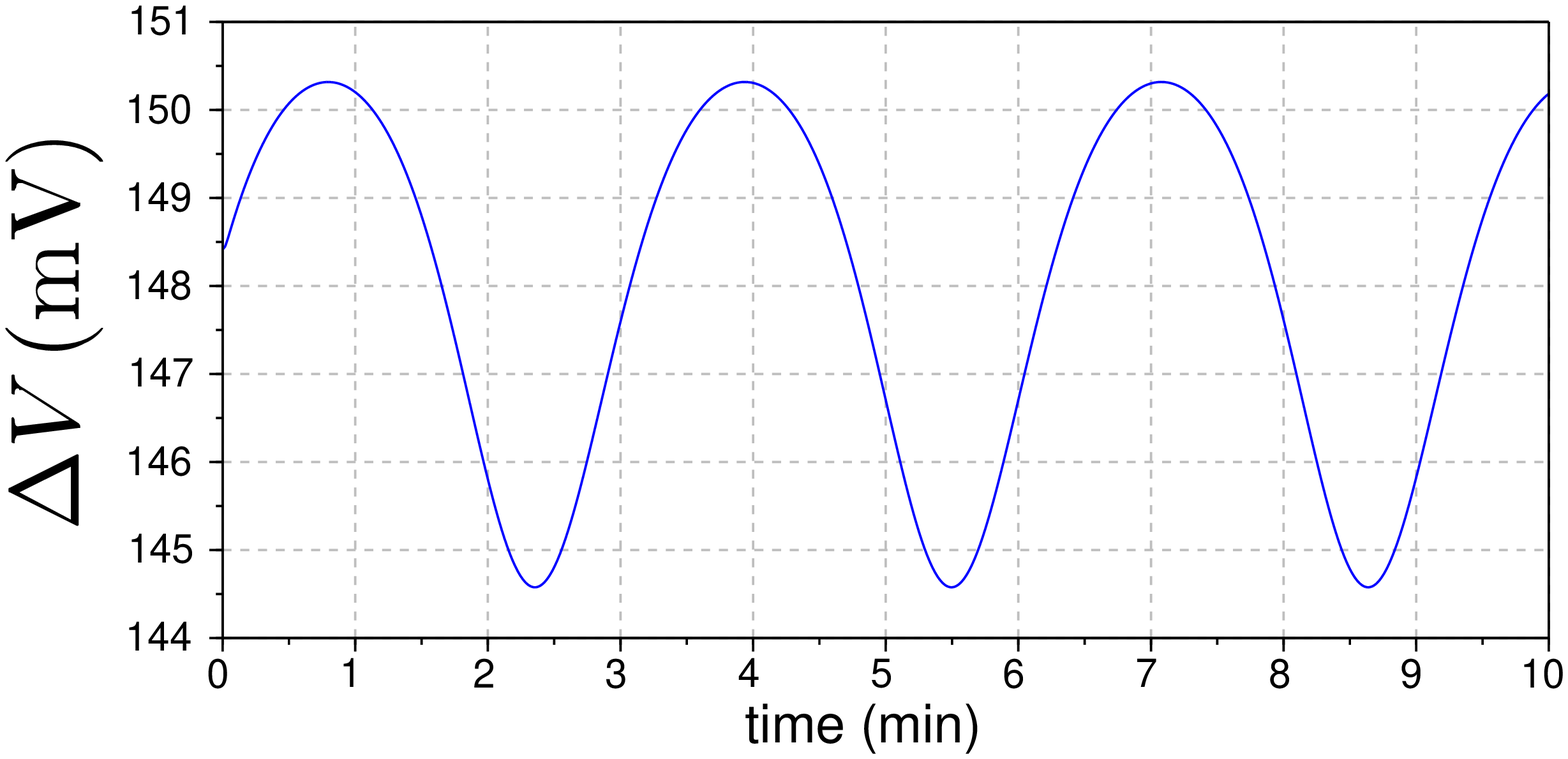}
  \includegraphics[width=0.5\linewidth]{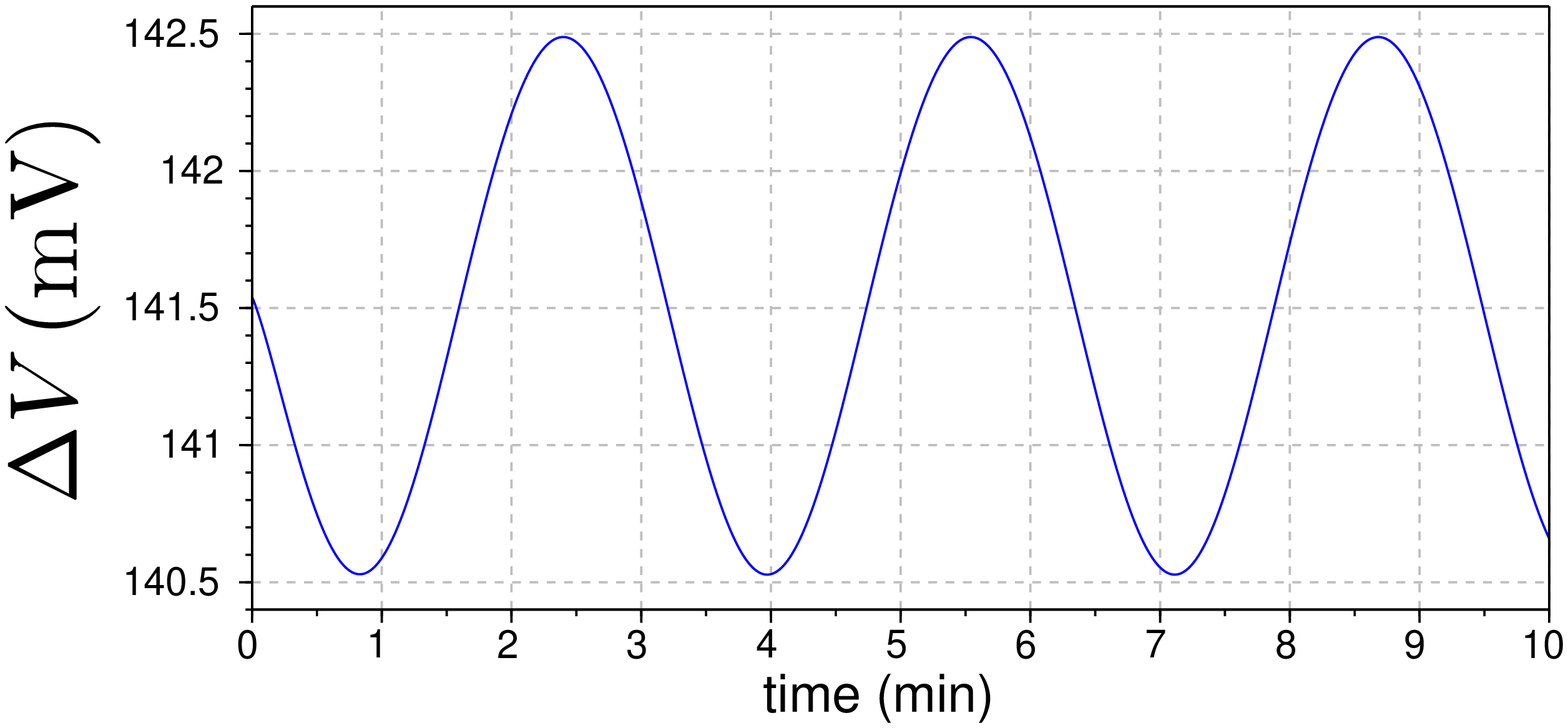}
\caption{Response of the equations (\ref{ee1})-(\ref{ee4}) to oscillatory inputs (\ref{osc}). Notice that for low ${\rm Ca}_{\rm c}$ concentrations (left), all the mitochondrial variables increases and decreases in synchrony with the variations of $u$. On the other hand, for high ${\rm Ca}_{\rm c}$ concentrations (right), the behavior of $x$, $y$ and $w$ is reversed. All the curves were evaluated for ${\rm FBP} = 0.5\,\mu$M. See the text for further details.}
\label{fig3}
\end{figure}

\begin{figure}[t!]
\begin{center}
\includegraphics[width=0.5\linewidth]{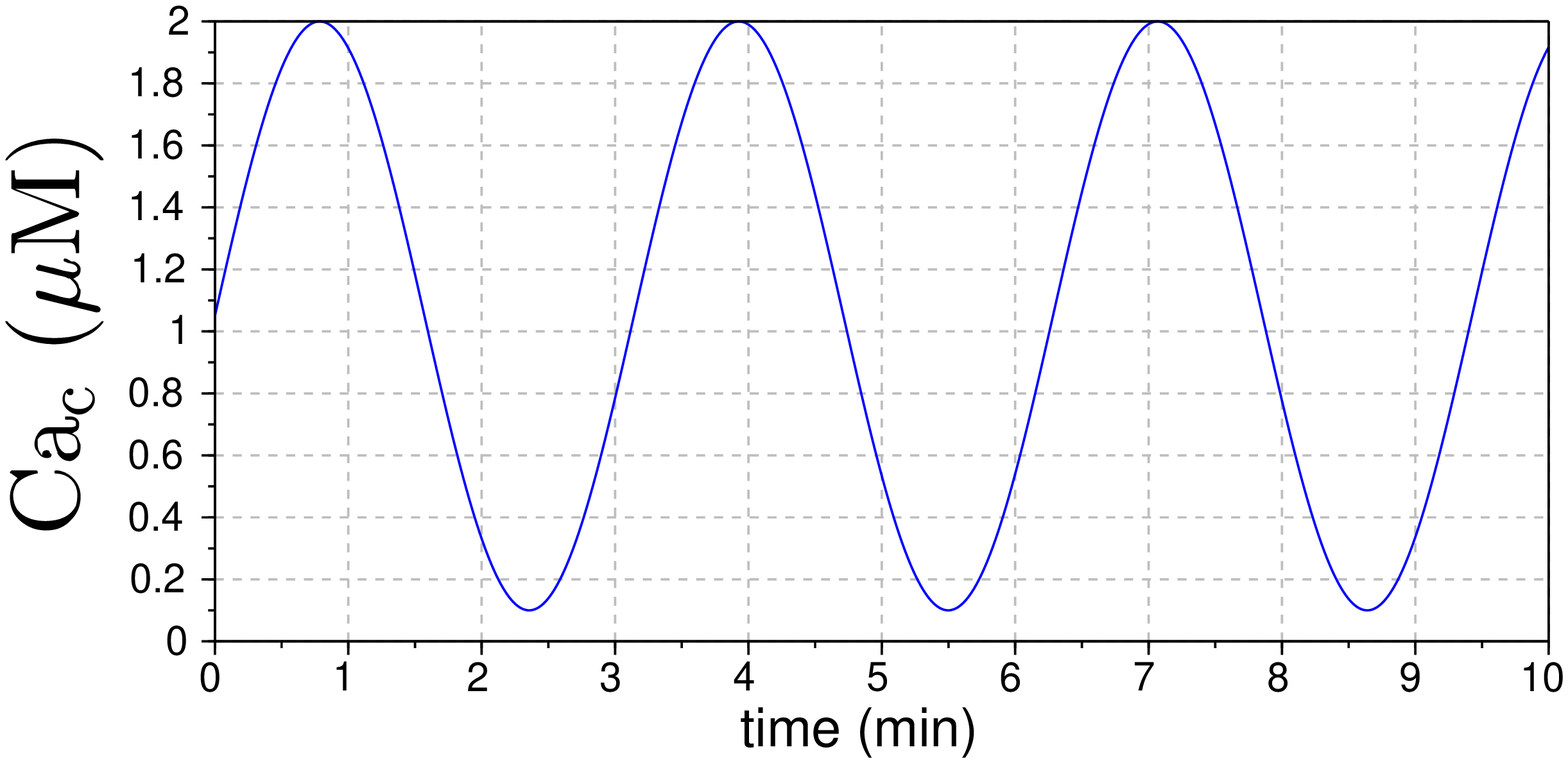}\\
\end{center}
  \includegraphics[width=0.5\linewidth]{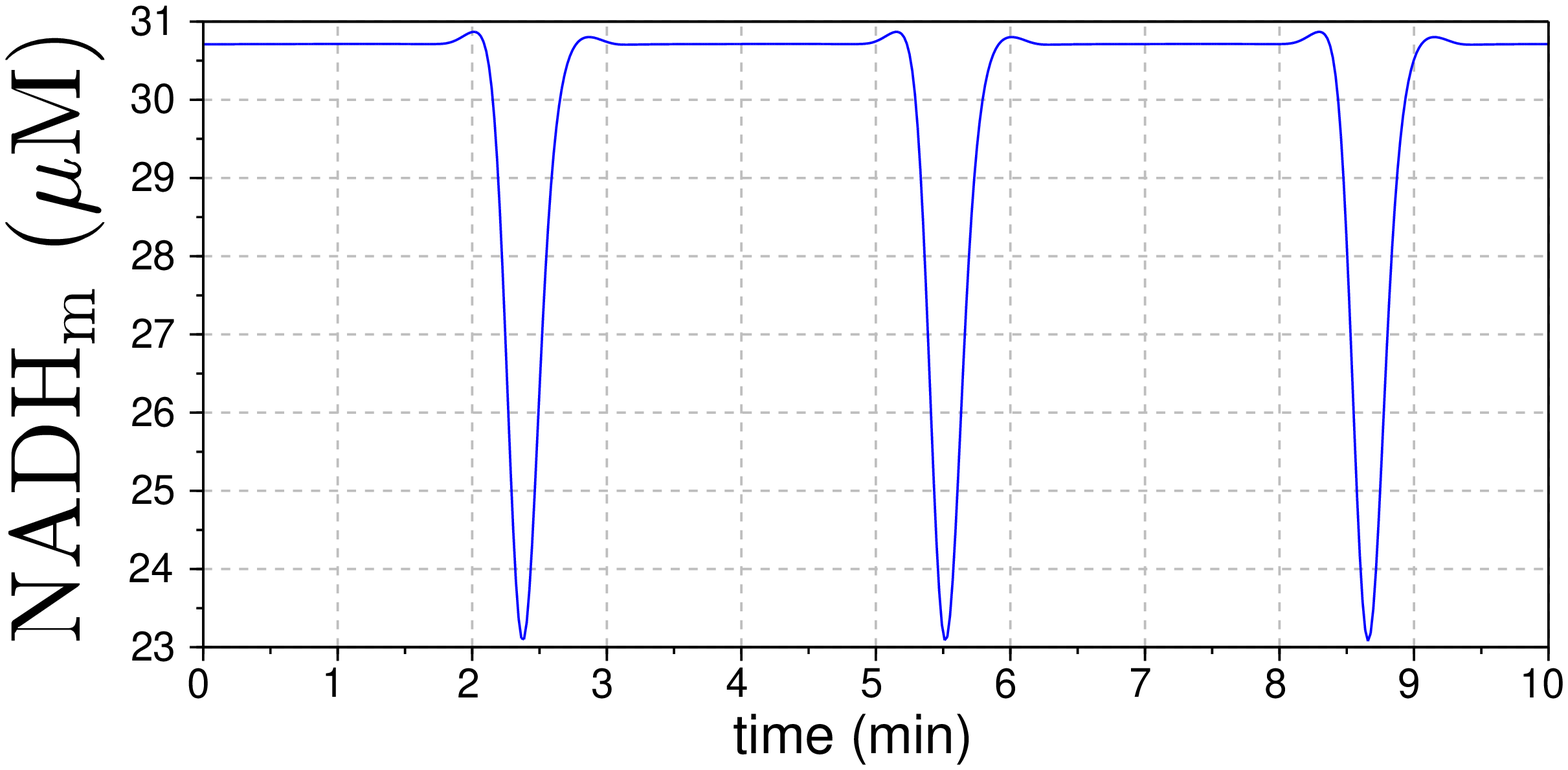}
  \includegraphics[width=0.5\linewidth]{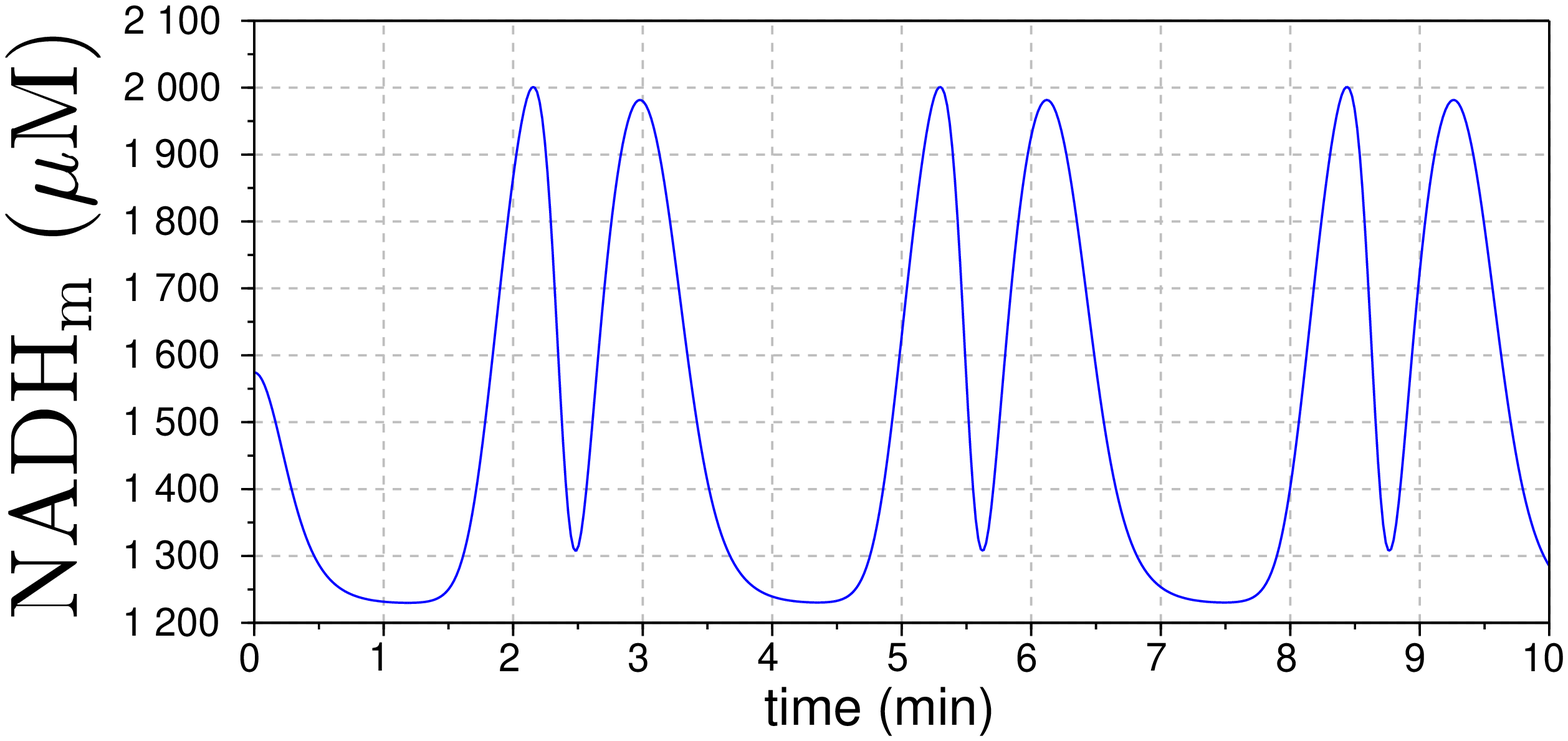}
  \includegraphics[width=0.5\linewidth]{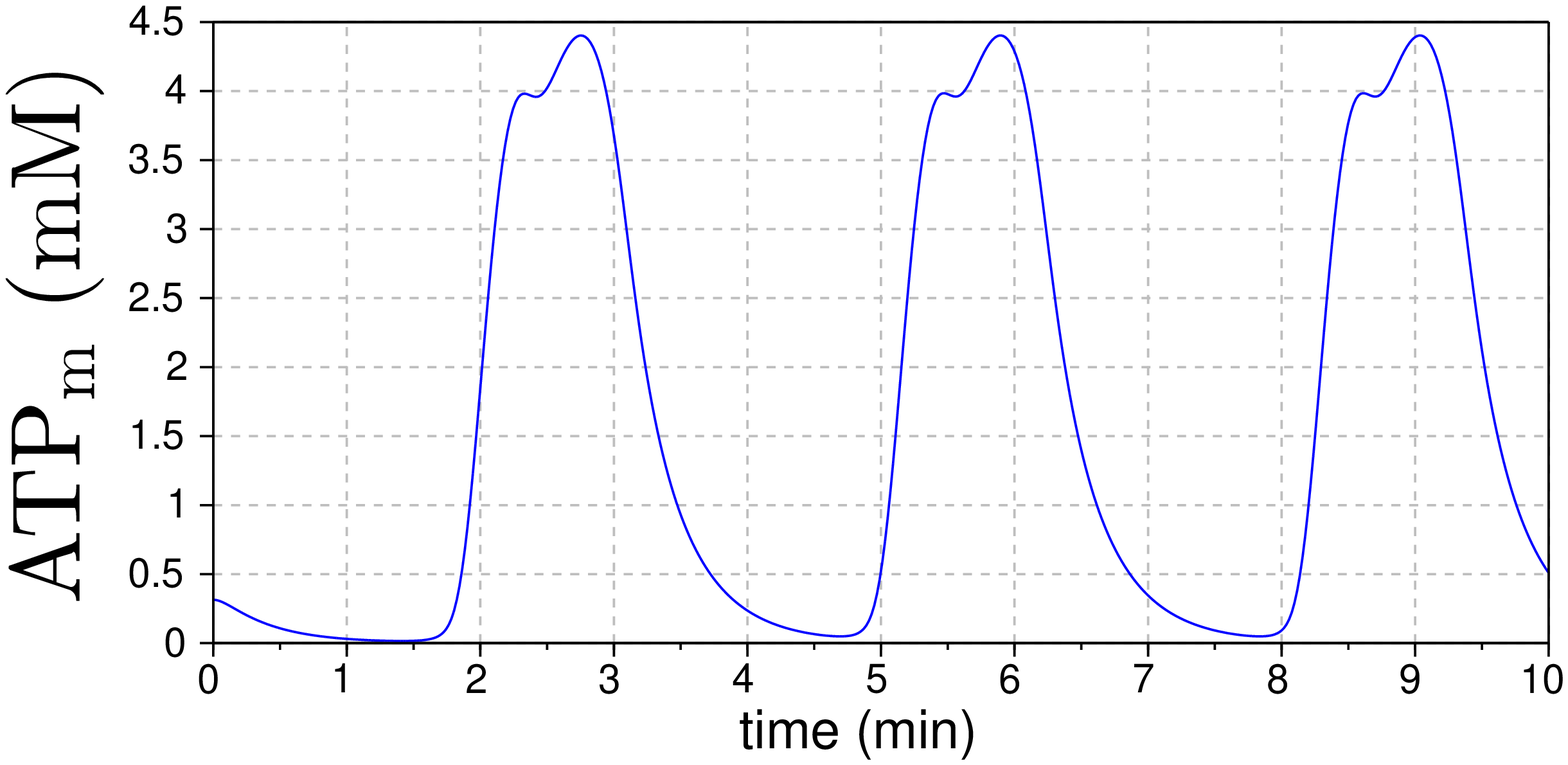}
  \includegraphics[width=0.5\linewidth]{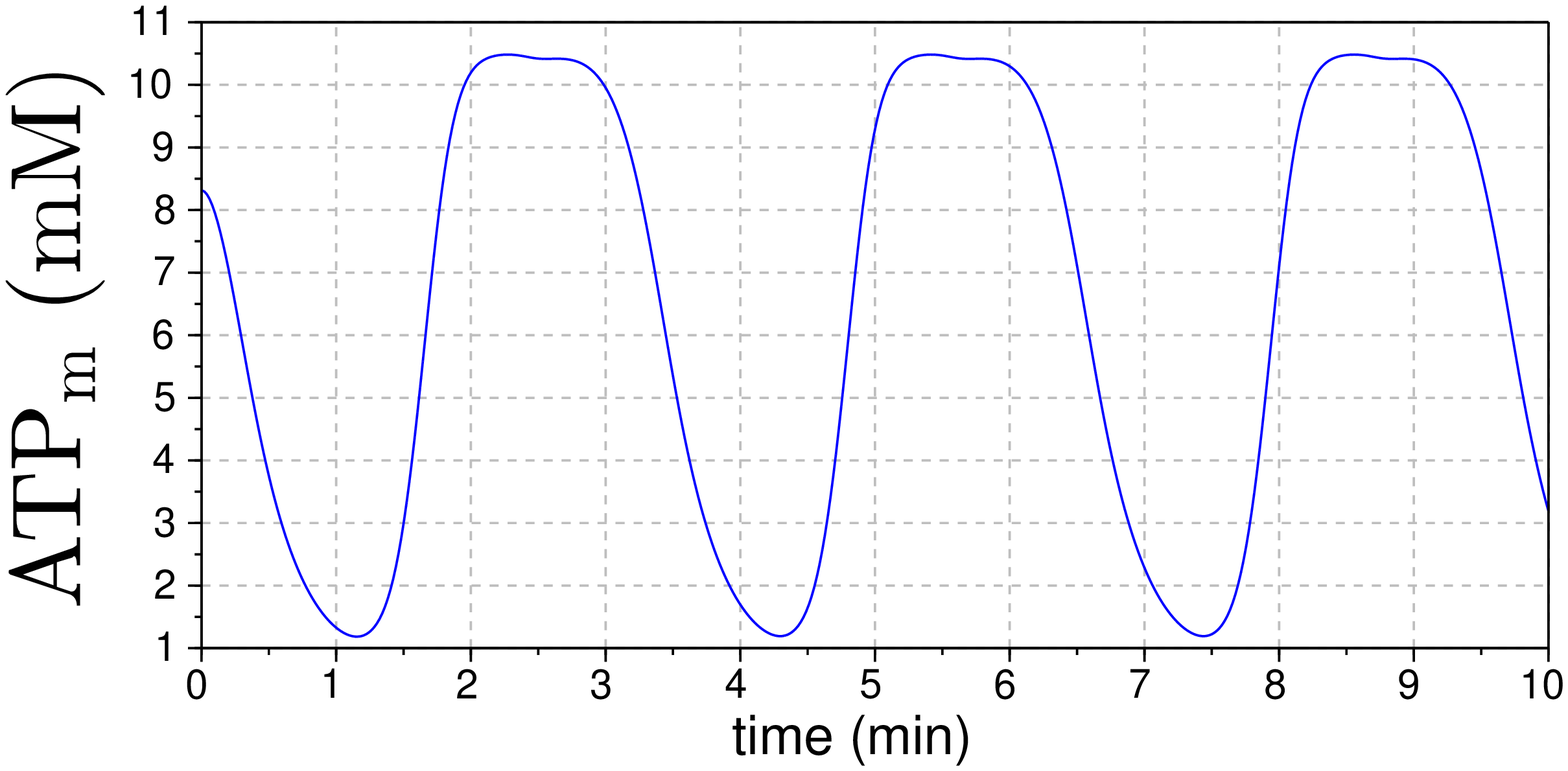}
  \includegraphics[width=0.5\linewidth]{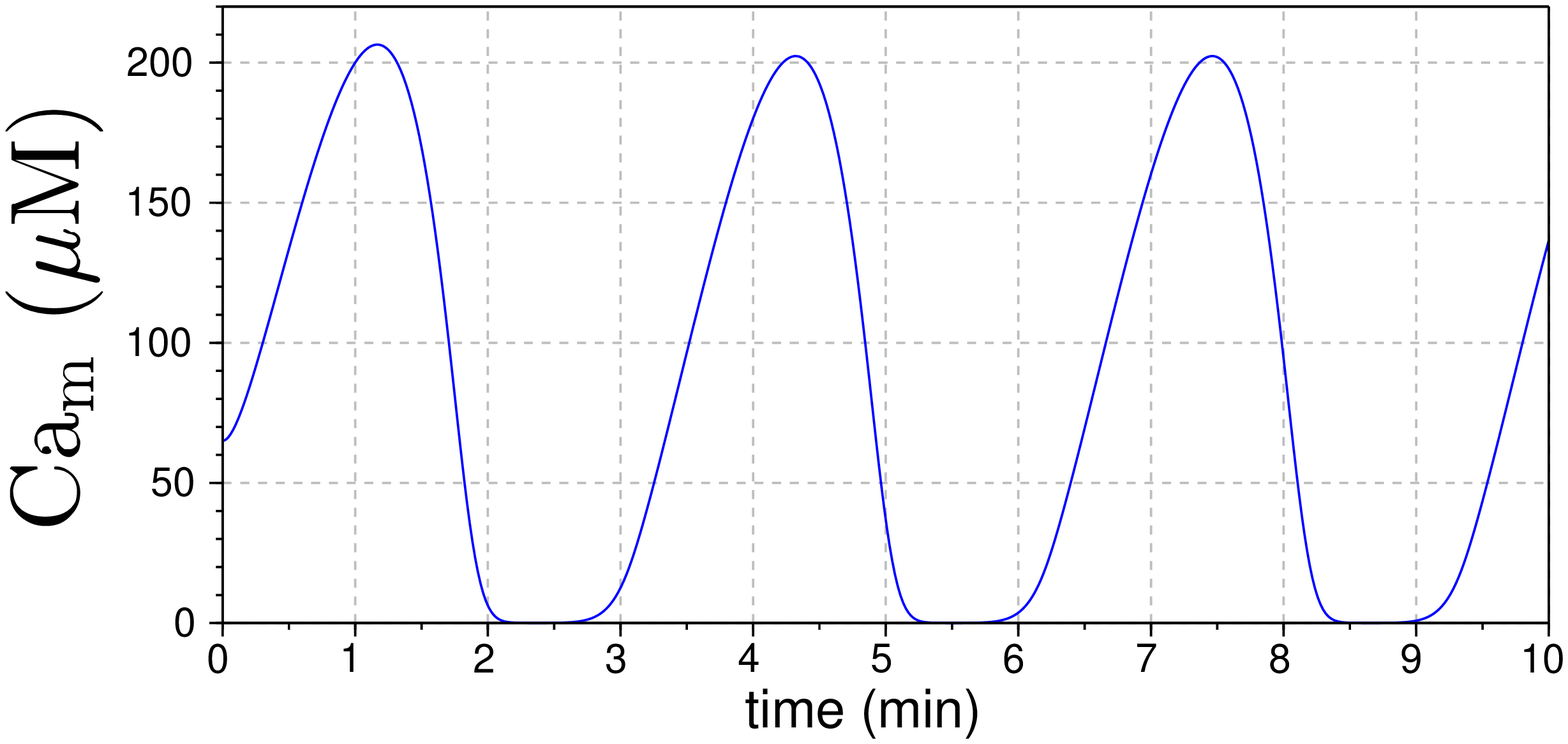}
  \includegraphics[width=0.5\linewidth]{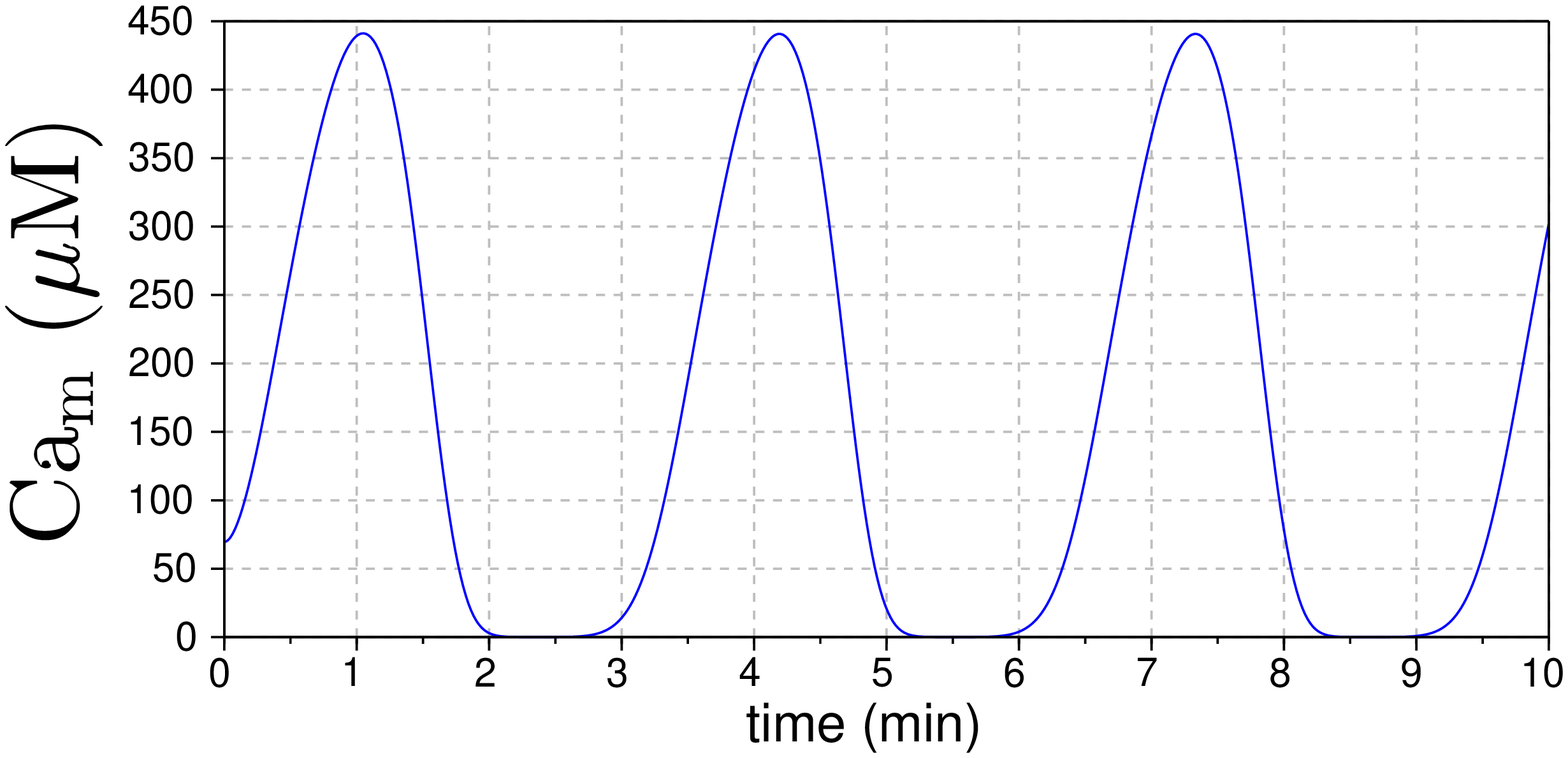}
  \includegraphics[width=0.5\linewidth]{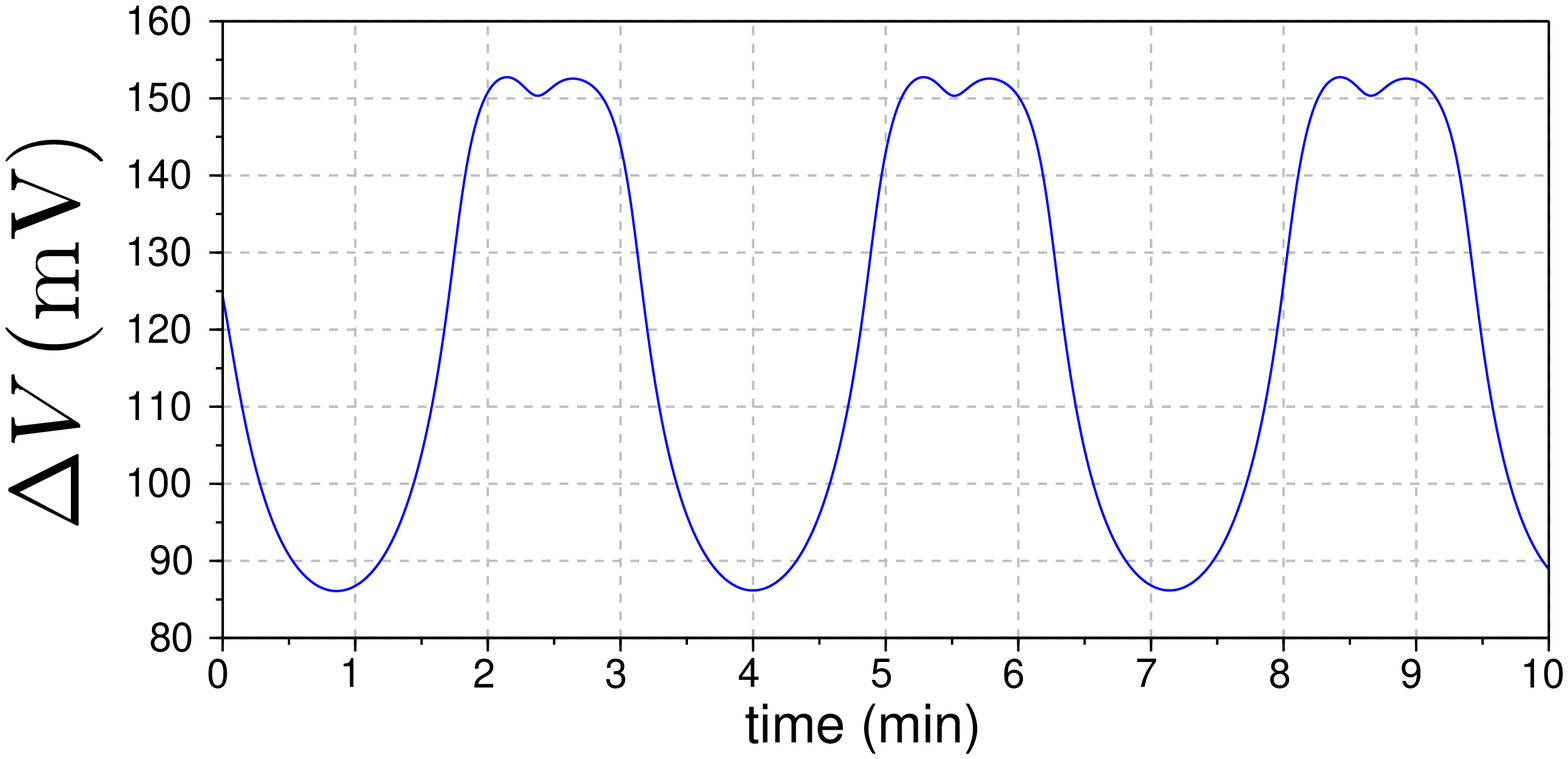}
  \includegraphics[width=0.5\linewidth]{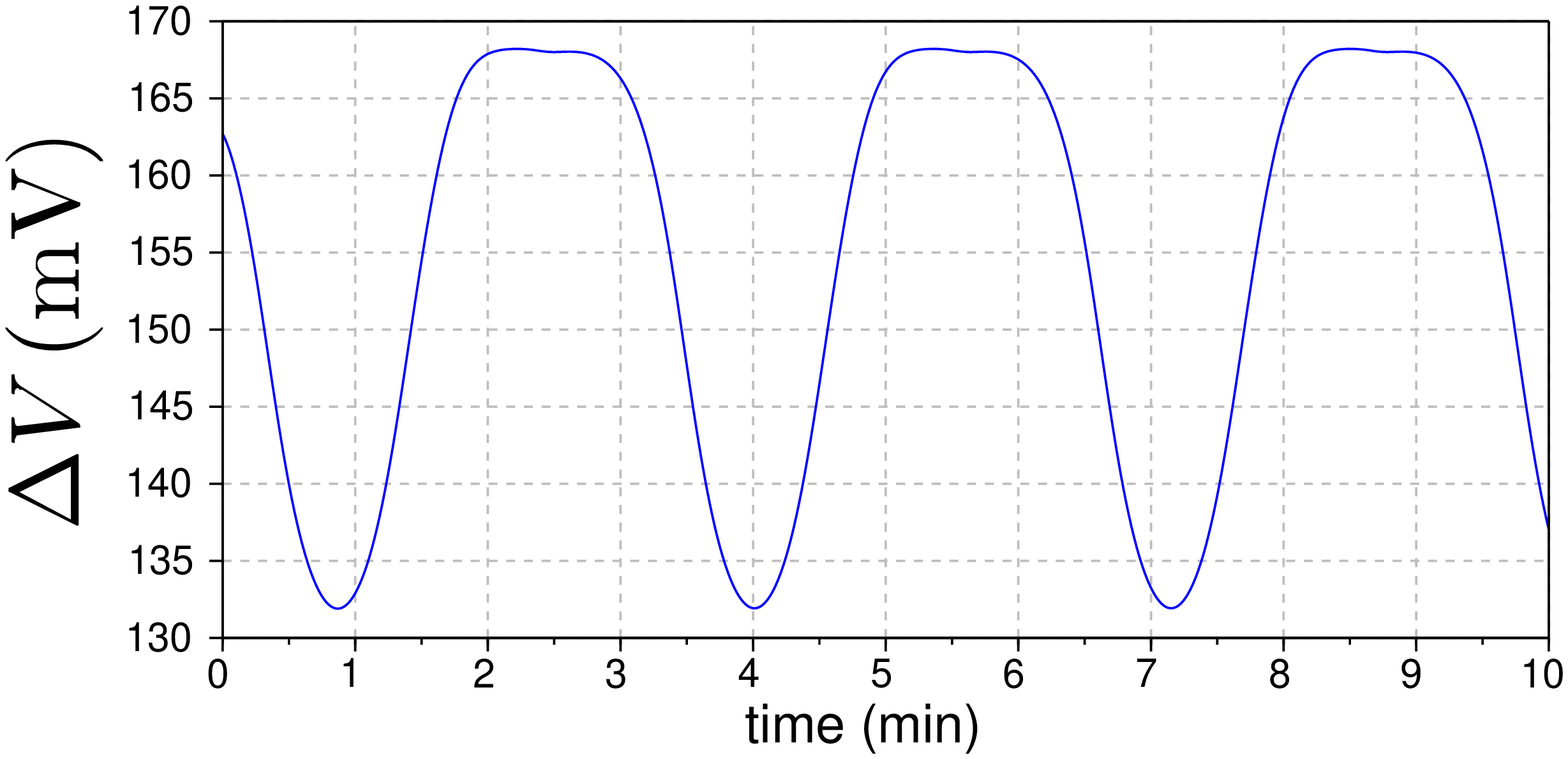}
\caption{Response of the equations (\ref{ee1})-(\ref{ee4}) to oscillatory inputs (\ref{osc}) for different concentrations of FBP. The system is submitted to the
oscillatory input corresponding to the top graphic. The left column is the response for ${\rm FBP} = 0.5\,\mu$M, while the right one corresponds to ${\rm FBP} = 10\, \mu$M.
Its clear that the ${\rm NADH}_{\rm m}$ concentration reverses its dynamical behavior for low and high concentrations of FBP.  See the text for further details.}
\label{fig4}
\end{figure}

Another important feature of the BPLS model is the reversion
of the dynamical behavior
 of some mitochondrial variables 
   in the presence of lower and higher concentration of  cytosolic  calcium
   and FBP. 
This behavior can be seen,  for instance, in Fig. (2b). After attaining its maximum,
the ATP production ($y_*$) tends to decrease for increasing  cytosolic  calcium
concentrations ($u_*$).
    Calcium saturation can be
   simulated, as described in \cite{bertram}, by setting $a_{1}=0$ in the expression for
$J_{\rm PDH}$ (\ref{JPDH}). The reversion
of the dynamical behavior of the other mitochondrial variables for higher ${\rm Ca}_{\rm c}$ concentrations  can also be inferred directly from Fig. \ref{fig2}, but
it is certainly better illustrated in Fig. \ref{fig3},
which depicts the solutions of (\ref{ee1})-(\ref{ee4}) for an oscillatory ${\rm Ca}_{\rm c}$ input of the form
\begin{equation}
\label{osc}
u(t) = u_0 + u_1\sin( {t}/{t_0}),
\end{equation}
with constant $v(t)$ 
and initial conditions $(x(0),y(0),z(0),w(0))$ given by the values of the fixed point 
corresponding to $u_*=u(0)$ and $v_*=v(0)$. As we will see, such a choice
of initial condition  is consistent with the adiabatic (stationary) regime we observe for sufficiently slow inputs (large periods $t_0$). For lower values of $u_0$ 
(low ${\rm Ca}_{\rm c}$ concentrations), all the mitochondrial variables increases and decreases in synchrony with the variations of $u$. On the other hand, for higher 
values of $u_0$, the dynamical behavior of $x$, $y$ and $w$ is reversed.
{\em i.e.}, they tend to decrease/increase while $u$ increases/decreases. This effect can be understood from the relation between $u_*$ and $z_*$ depicted in Fig. (2c). The value of $z_*$ tends to increase rapidly when $u_*$ increases and, for large values of $z$, the dependence of the expression for $J_{\rm PDH}$ (\ref{JPDH}) on $z$ saturates and  becomes equivalent to setting $a_1=0$. Without
the $z$ suppression term in $J_{\rm PDH}$, the dynamical behavior of the variables 
$x$, $y$, and $w$ is reversed, as it was pointed out in the original BPLS analysis. Low variations  of $v$ (FBP) do not change qualitatively this dynamical behavior. However, the situation changes for large concentrations of FBP. As described in
\cite{bertram}, for low concentrations of FBP, the ${\rm NADH}_{\rm m}$ concentration reacts to a sudden rising of 
${\rm Ca}_{\rm c}$  with an upward teeth, while for high concentrations of FBP such behavior is reversed, {\em i.e.}, ${\rm NADH}_{\rm m}$  concentration exhibits a downward teeth if ${\rm Ca}_{\rm c}$ increases. This situation is analyzed and depicted in Fig. (\ref{fig4}).
\begin{figure}[h!]
  \includegraphics[width=0.5\linewidth]{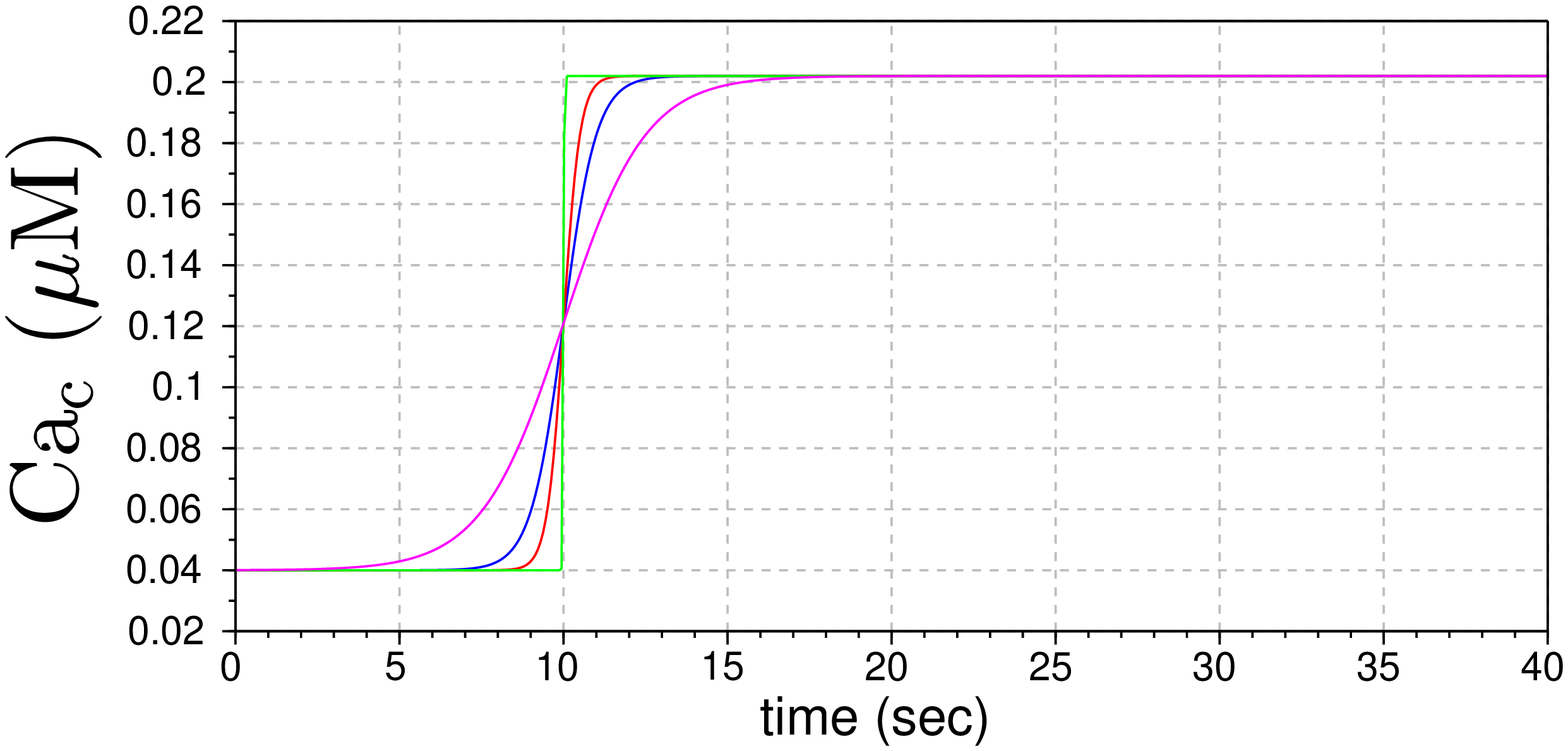}
  \includegraphics[width=0.5\linewidth]{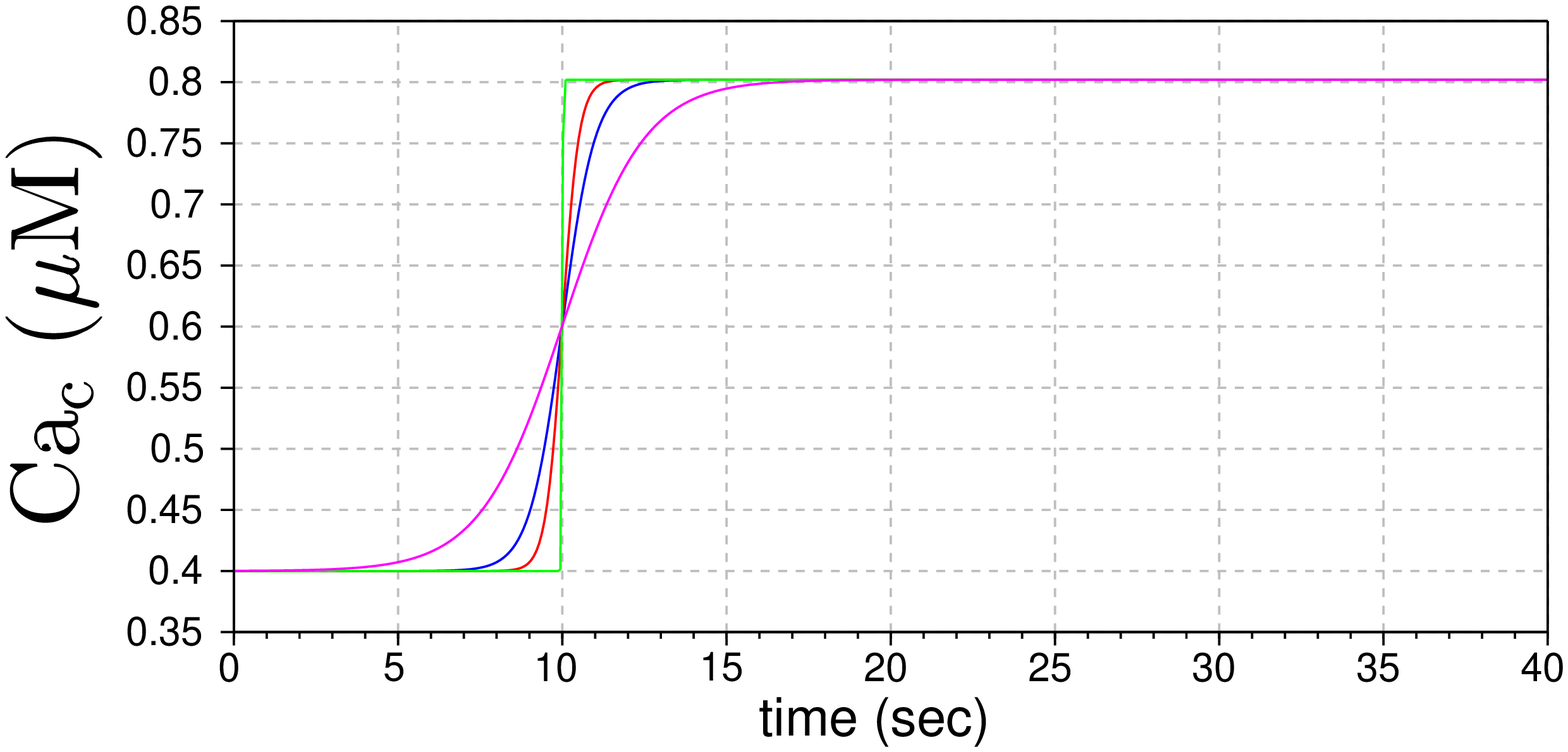}
  \includegraphics[width=0.5\linewidth]{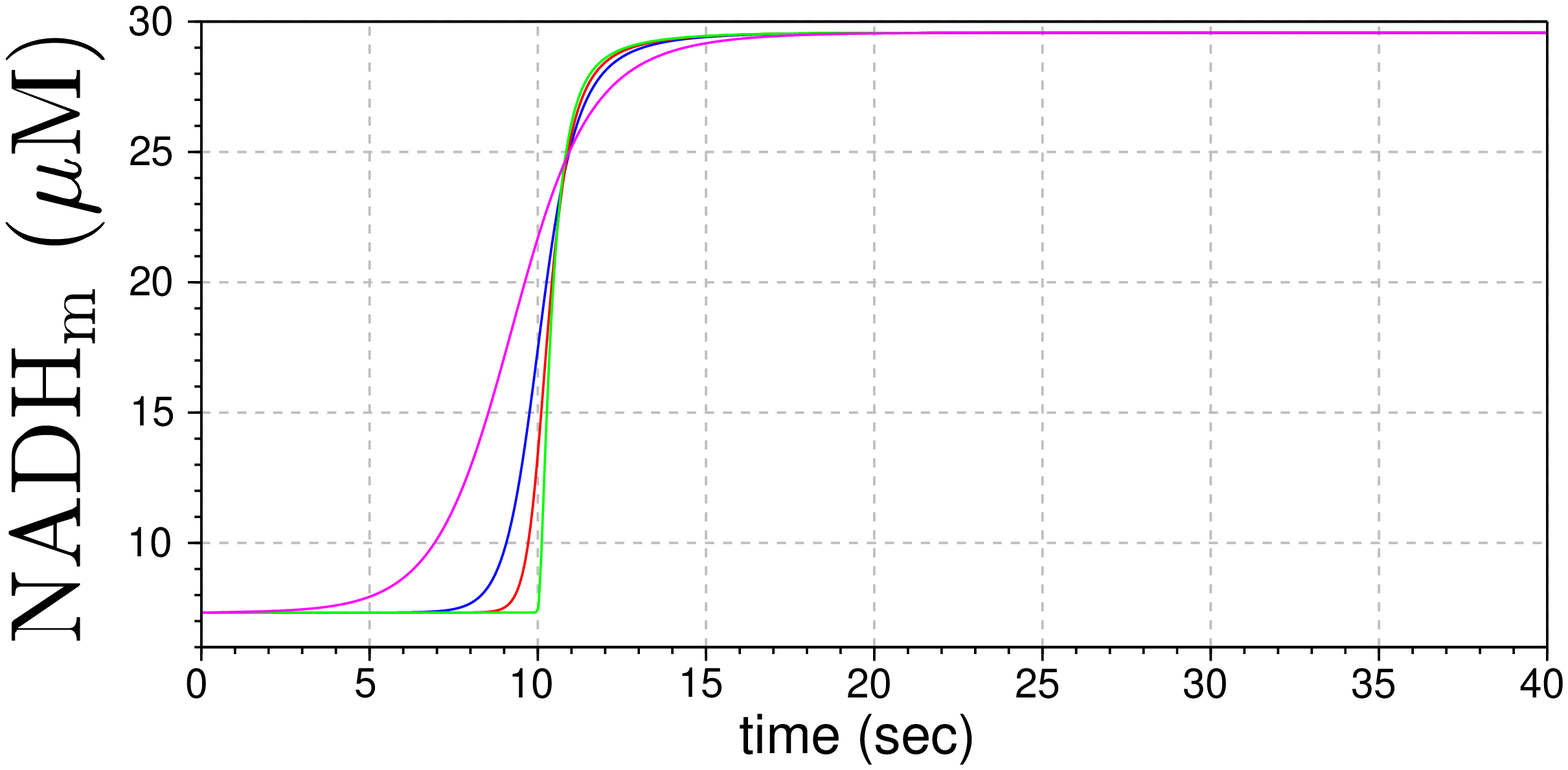}
  \includegraphics[width=0.5\linewidth]{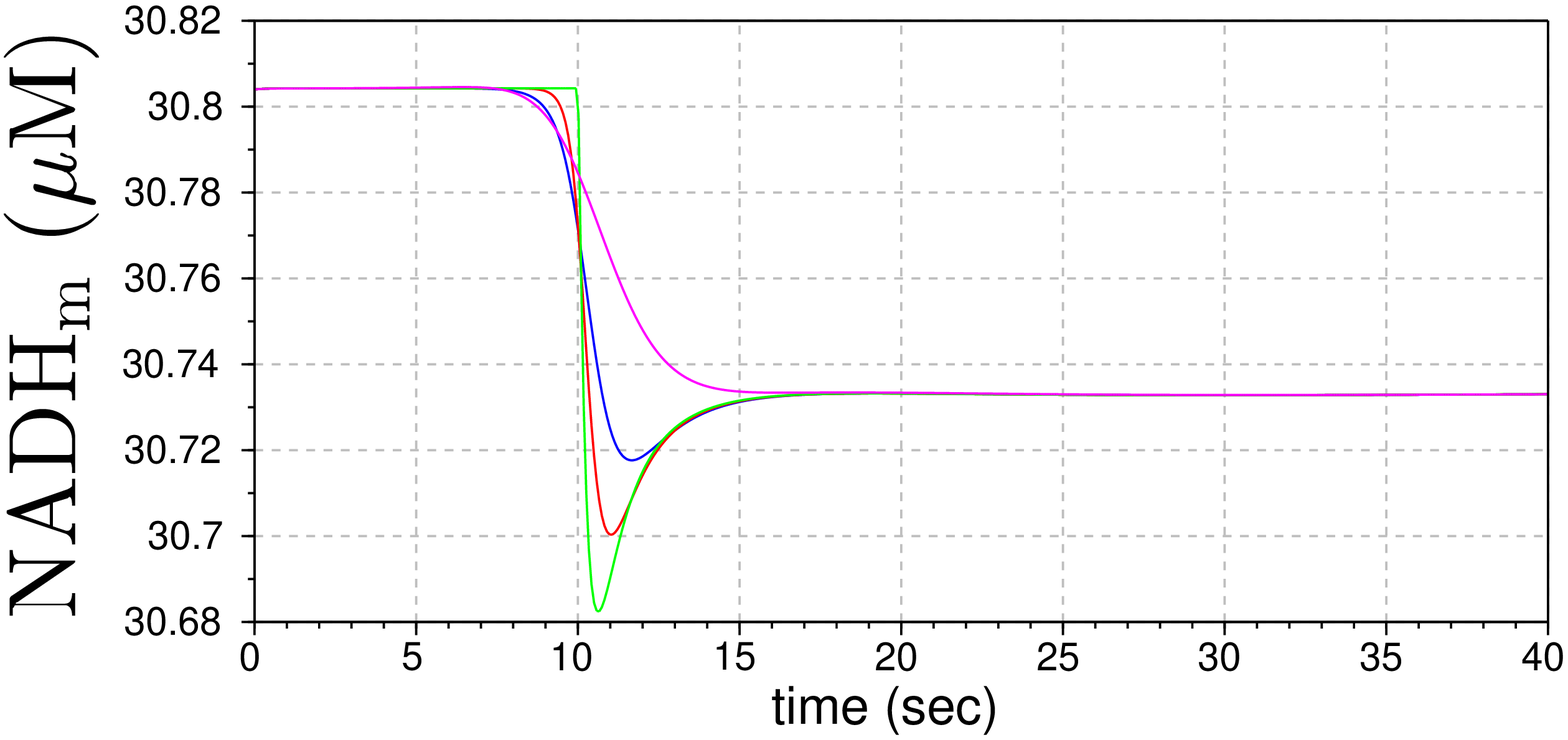}
  \includegraphics[width=0.5\linewidth]{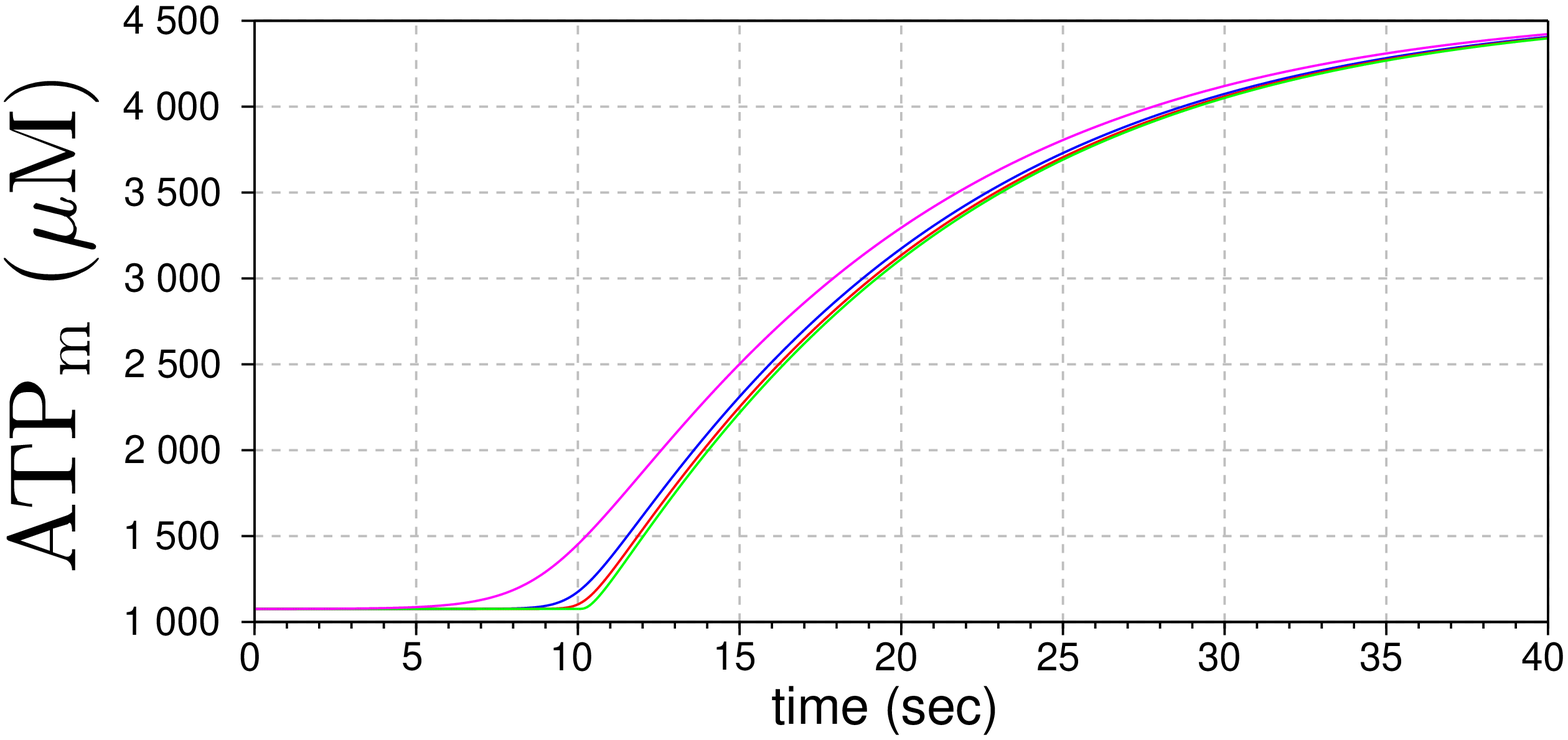}
  \includegraphics[width=0.5\linewidth]{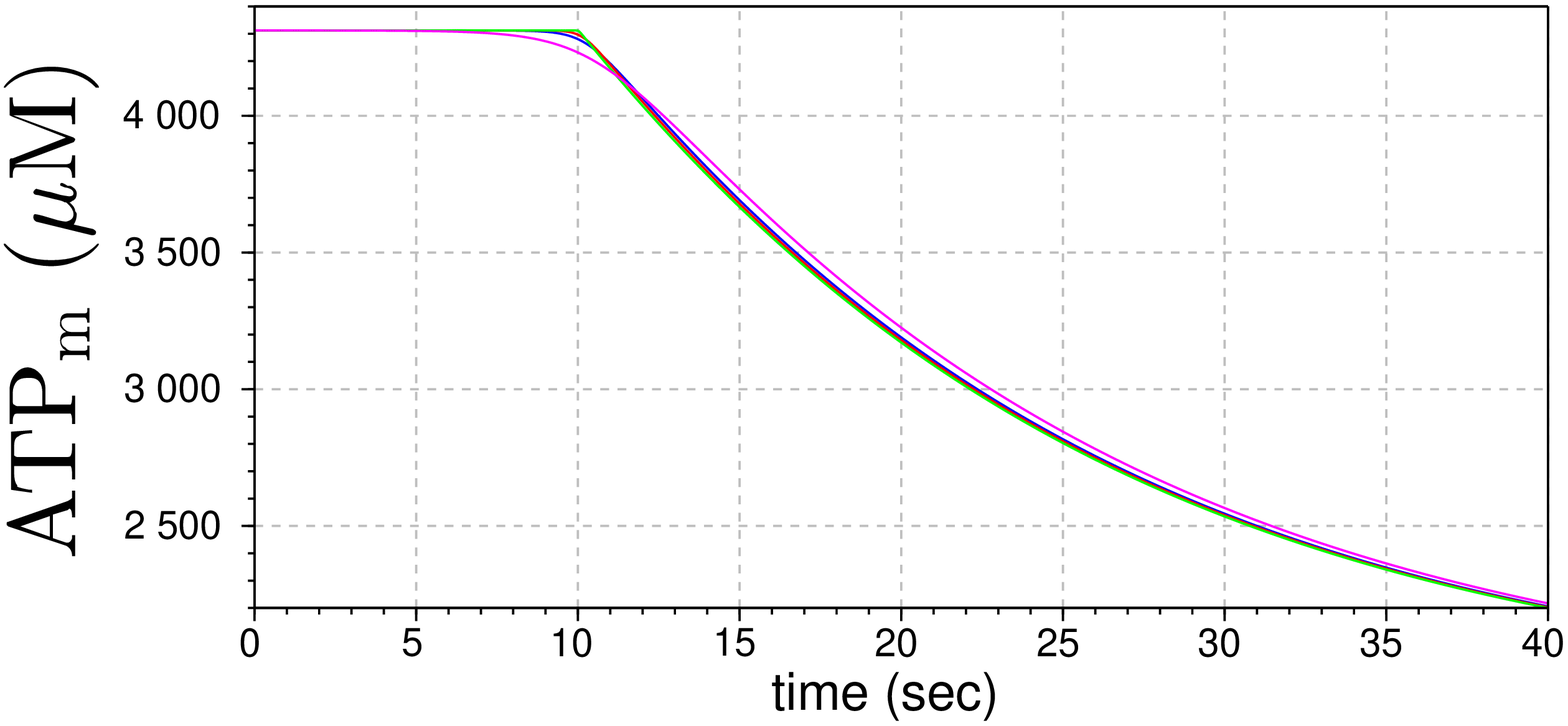}
  \includegraphics[width=0.5\linewidth]{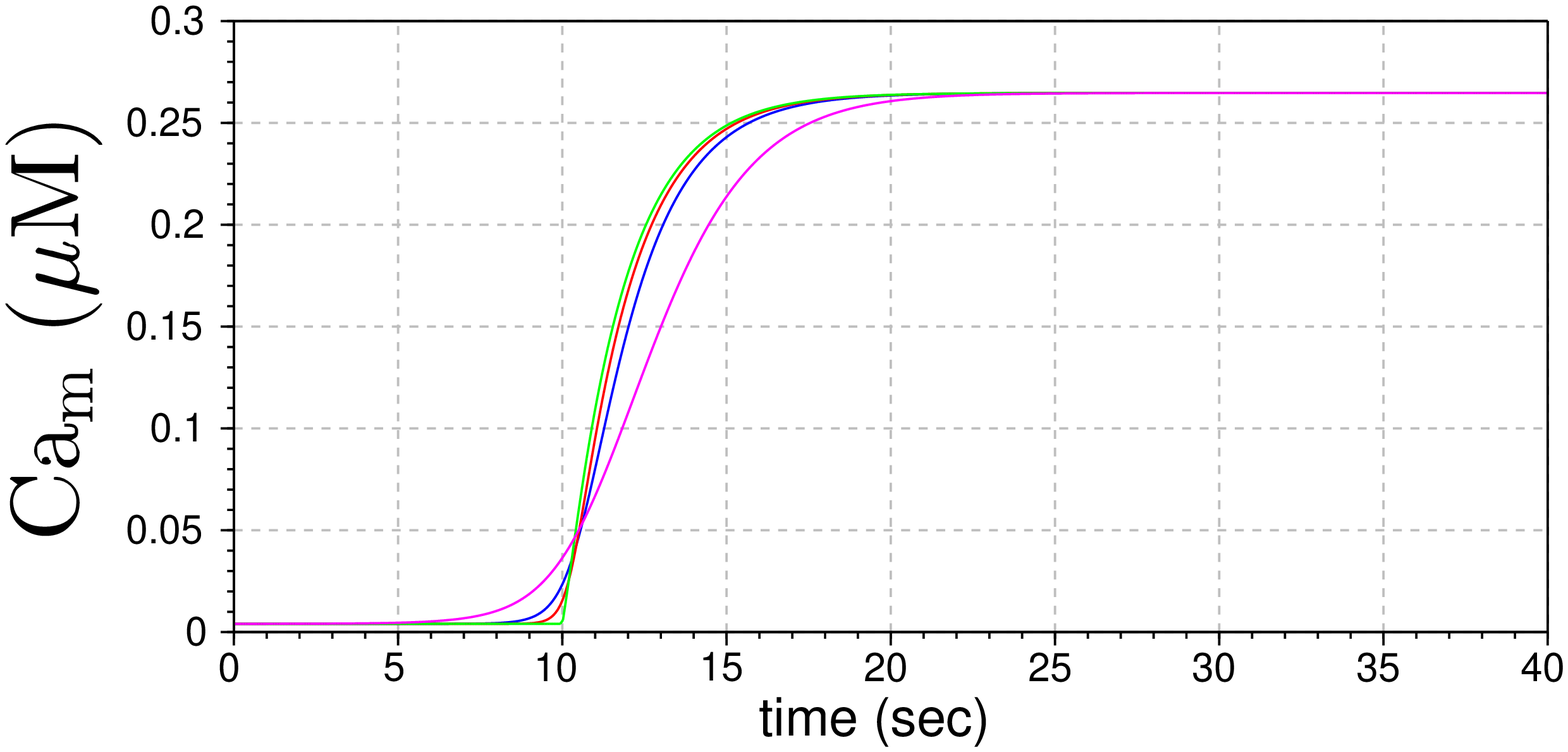}
  \includegraphics[width=0.5\linewidth]{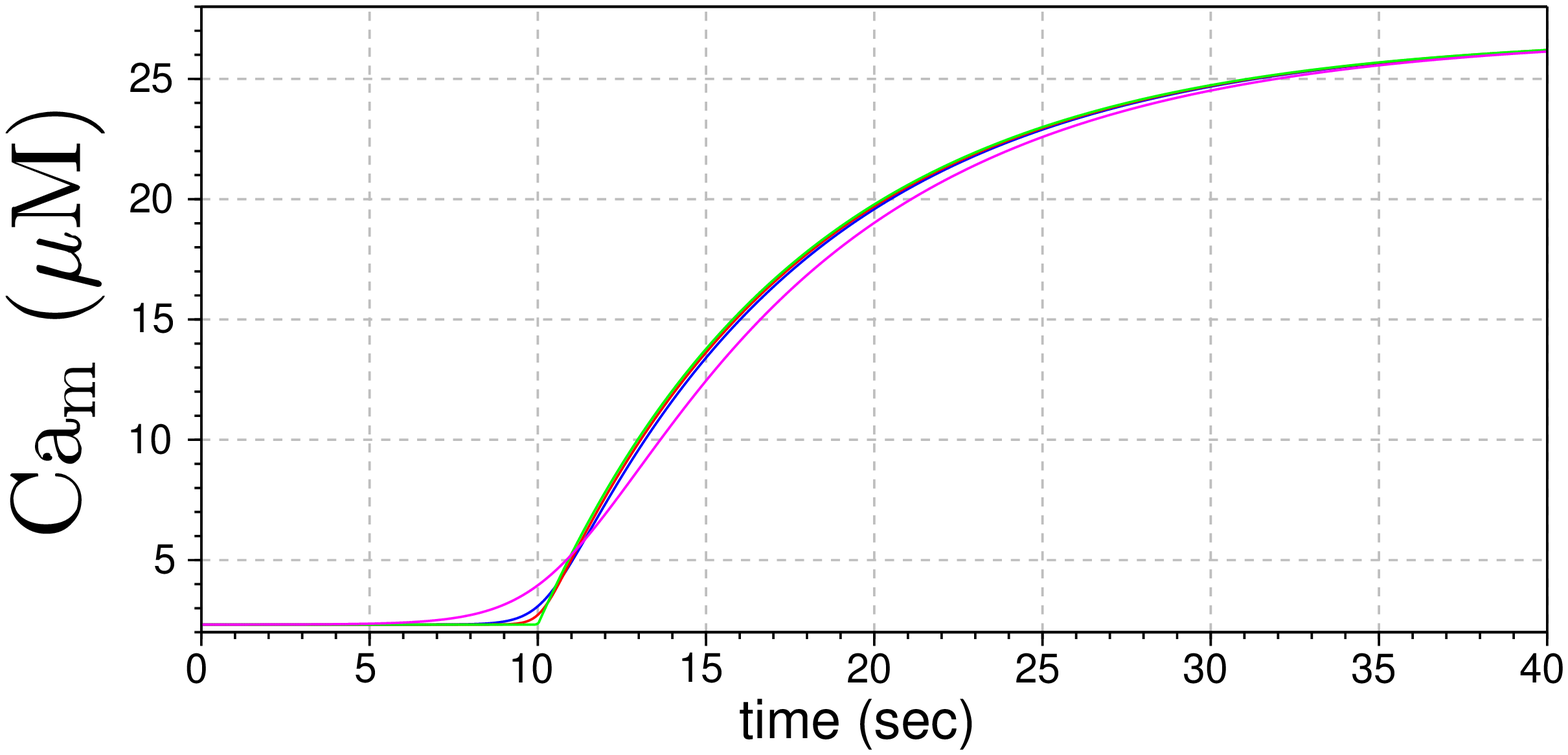}
  \includegraphics[width=0.5\linewidth]{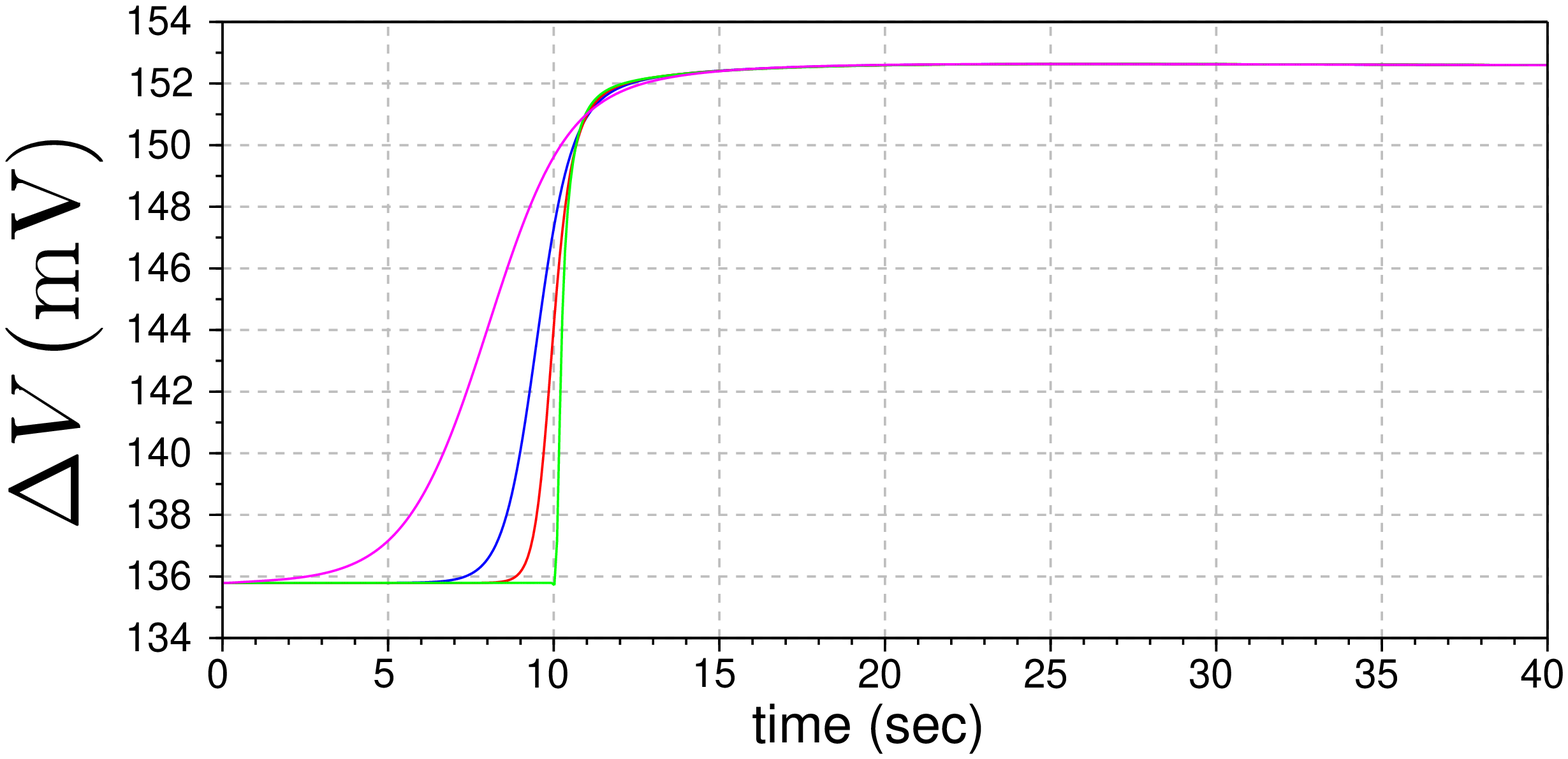}
  \includegraphics[width=0.5\linewidth]{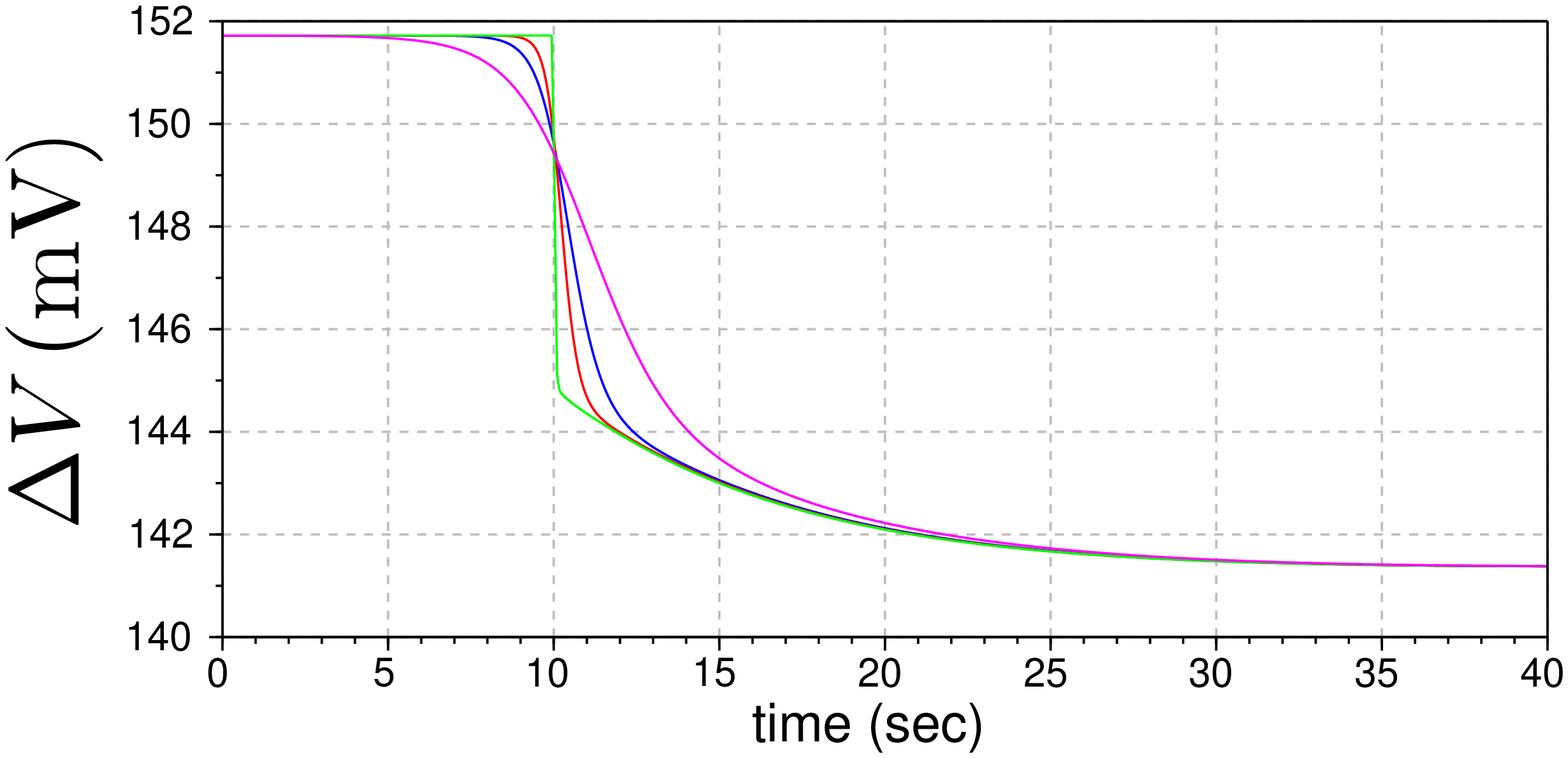}
\caption{Response of the equations (\ref{ee1})-(\ref{ee4}) to step-like inputs (\ref{tanh}). The   magenta, blue, red, and green curves correspond, respectively, to
$t_0= 2.5, 1, 0.5,$ and $0.02\,$seconds. 
The inertia of the system increases considerably for higher values
of $u$, see the text for further details. The curves were evaluated for
${\rm FBP} = 0.5\,\mu$M.}
\label{fig5}
\end{figure}

The oscillatory excitations used in the examples depicted in Fig. (\ref{fig3}) and 
(\ref{fig4}) have period $t_0 = 3\,$min. For inputs varying over a time scale of minutes, the system evolves adiabatically in a good approximation, {\em i.e.}, the instantaneous solution 
$(x(t),y(t),z(t),w(t))$ is well approximated by the fixed point $(x_*,y_*,z_*,w_*)$ corresponding to $u_*=u(t)$ and
$v_*=v(t)$. In other words, for slowly varying inputs, the solutions of the system are confined to the fixed-point surfaces depicted in Fig. \ref{fig2}. Of course, one expects a breakdown of this adiabatic  behavior for rapidly varying inputs. Non-stationary effects must appear for inputs varying with a characteristic time smaller than a certain critical value. In order to study non-stationary effects in our model, we consider the response of the system for inputs of the type
\begin{equation}
\label{tanh}
u(t) = u_0 + u_1\tanh\left( \frac{t - t_1}{t_0} \right),
\end{equation}
for different values of $t_0$. This situation is depicted in Fig. \ref{fig5} for some values of $u_0$ and $u_1$ and for $t_0= 2.5, 1, 0.5,$ and $0.02\,$seconds. It is clear that for lower values of $u$ (${\rm Ca}_{\rm c}$), approximately 10 seconds are enough to assure that  ${\rm NADH}_{\rm m}$  concentration and $\Delta V$
 reaches their values corresponding to the adiabatic regime, which in this case corresponds to
the homeostasis. As we have already noticed,
 the variables $y$ (ATP) 
and $z$ (${\rm Ca}_{\rm m}$) are the slowest ones to attain their respective stationary regimes. For lower values of  f $u$ (${\rm Ca}_{\rm c}$), they spend approximately 40 seconds to stabilize.  
 Increasing the values of $u$ implies the increasing of such ``relaxation''
times, {\em i.e.}, a larger time is necessary to attain homeostasis. The second column in Fig. \ref{fig5} corresponds to a situation with $u\in[2,4]$, for which  almost 30 seconds are necessary to assure the attainment of the stationary regime for the rapid variable
$w$, whereas the slow one will need a few minutes. The inertia of the system, hence, increases considerably for  higher concentrations of 
cytosolic calcium.

Higher concentrations of fructose 1,6-bisphosphate FBP also imply an increasing of
the inertia of the system. 
\begin{figure}[t!]
\begin{center}
\includegraphics[width=0.5\linewidth]{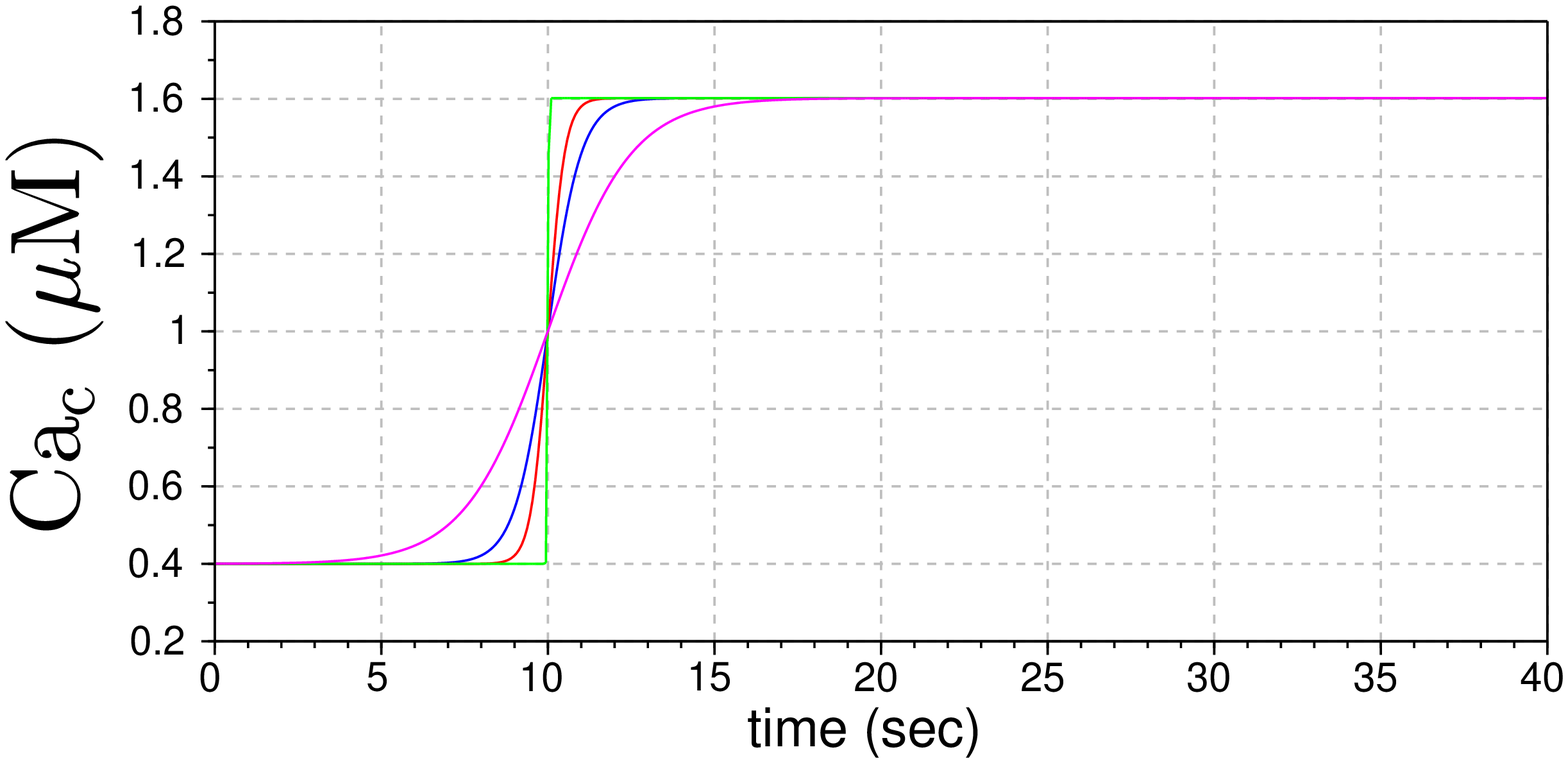}\\
\end{center}
  \includegraphics[width=0.5\linewidth]{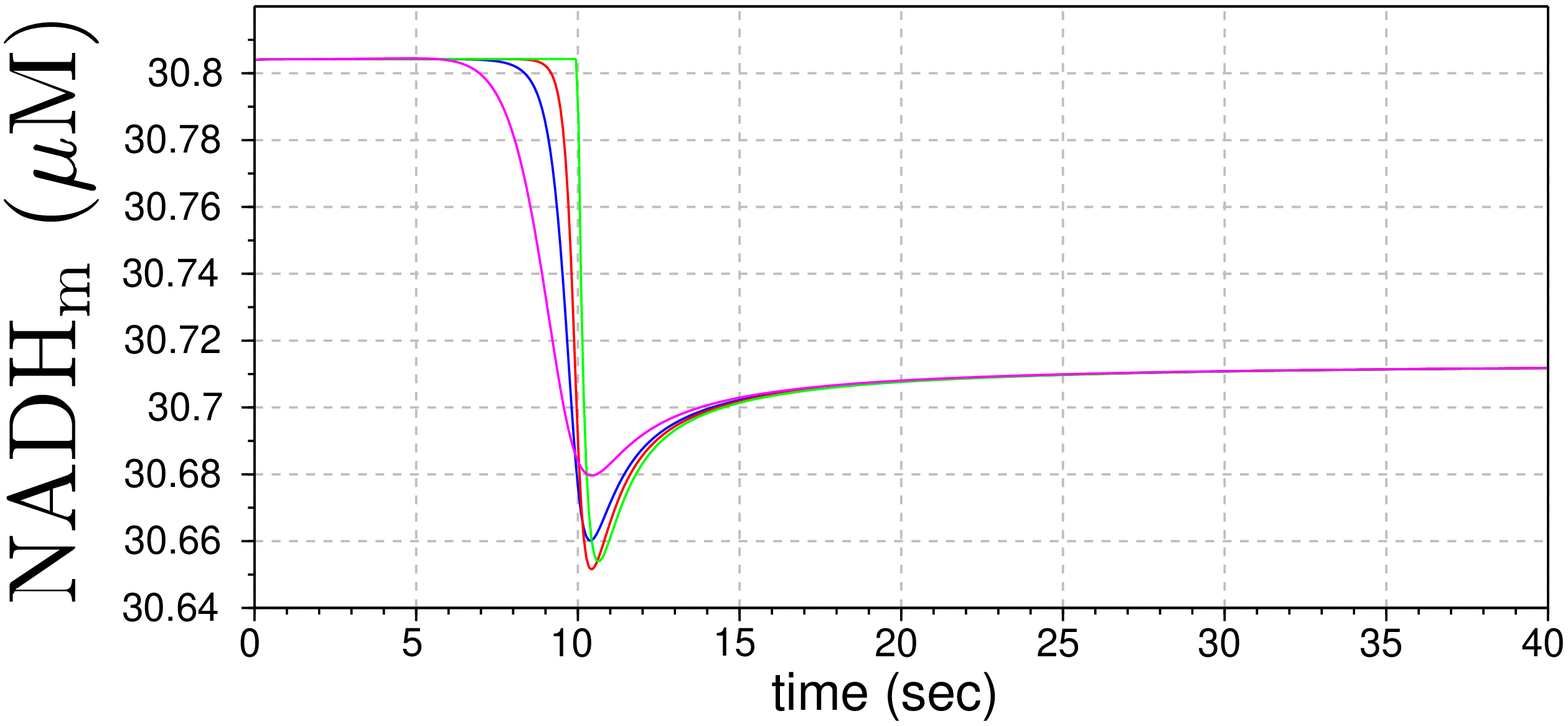}
  \includegraphics[width=0.5\linewidth]{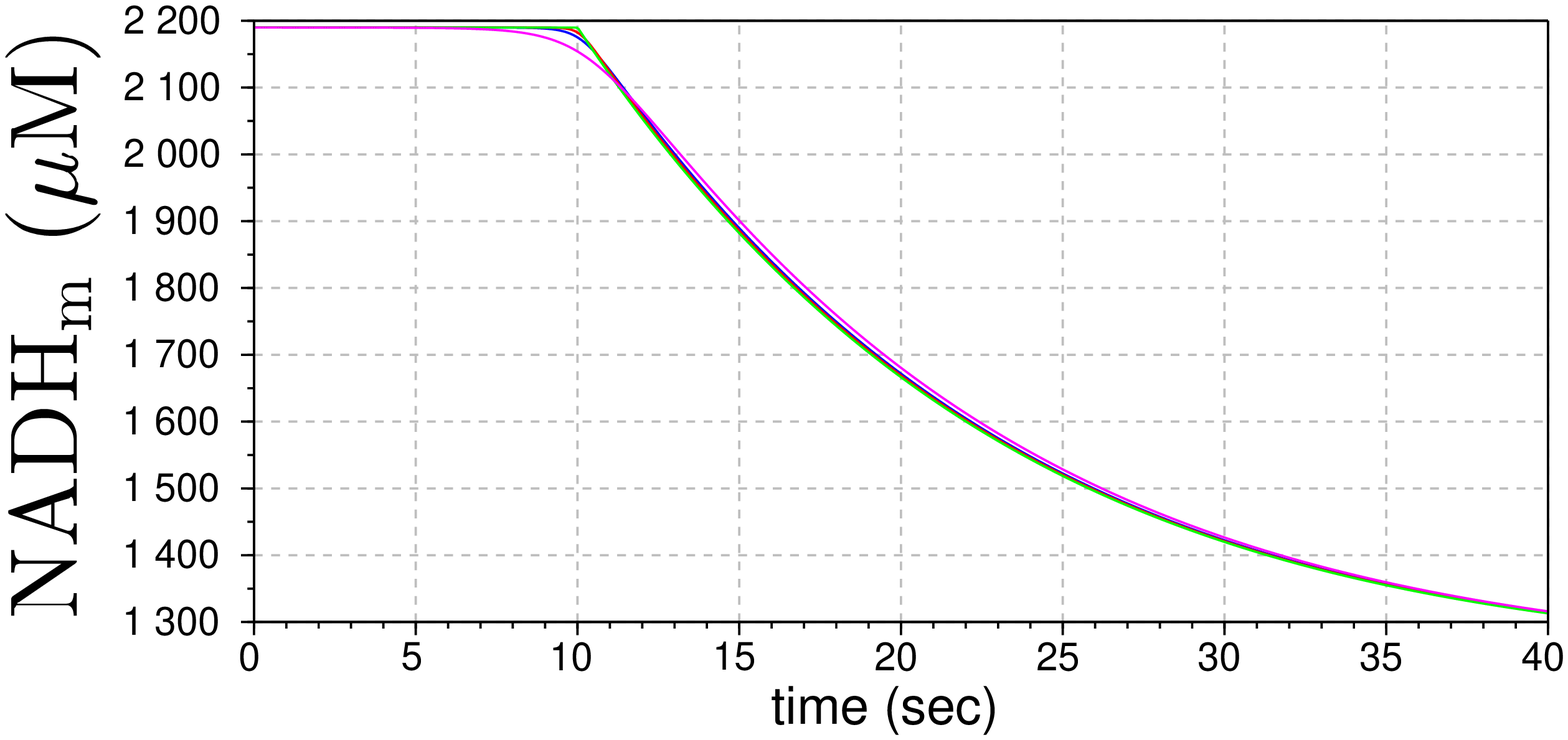}
  \includegraphics[width=0.5\linewidth]{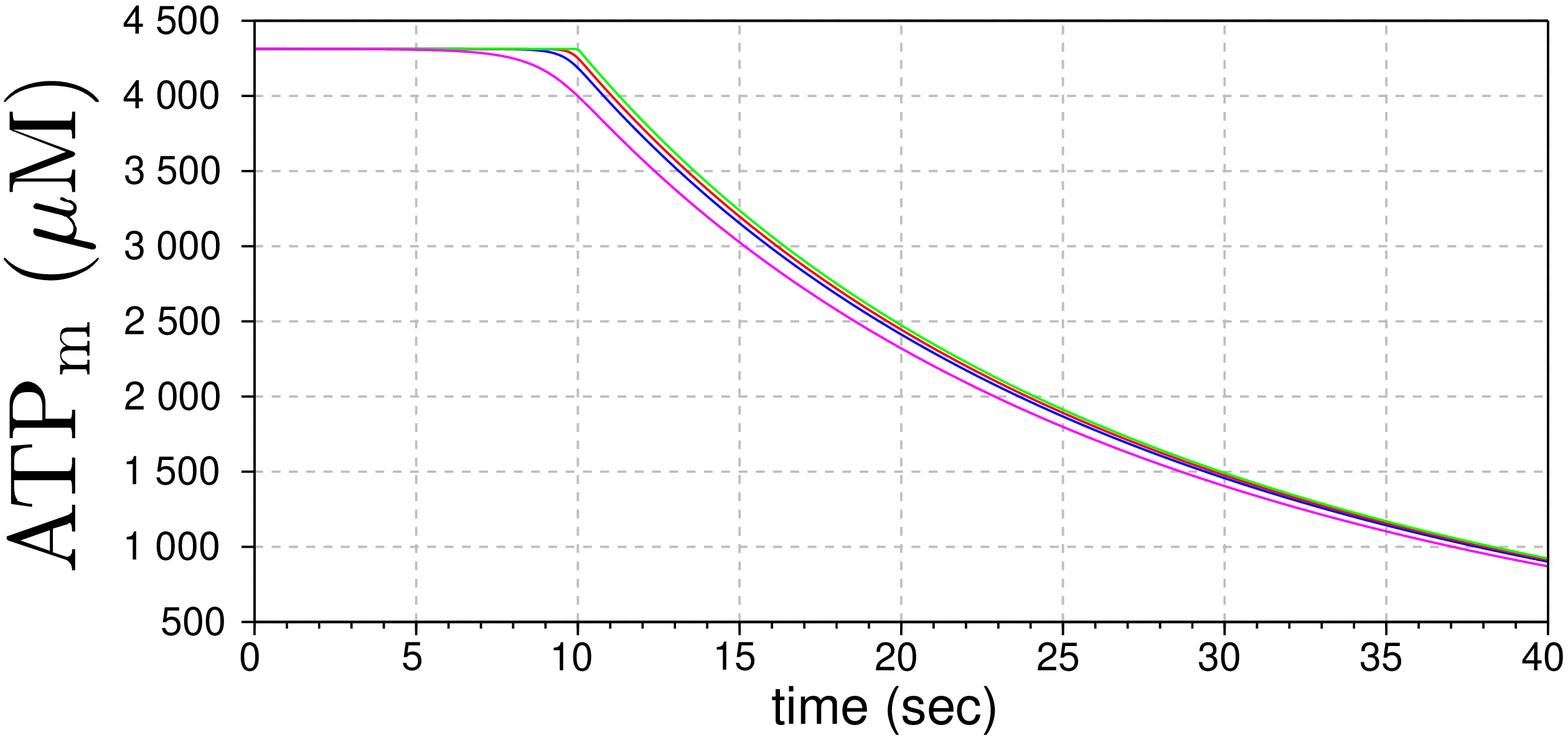}
  \includegraphics[width=0.5\linewidth]{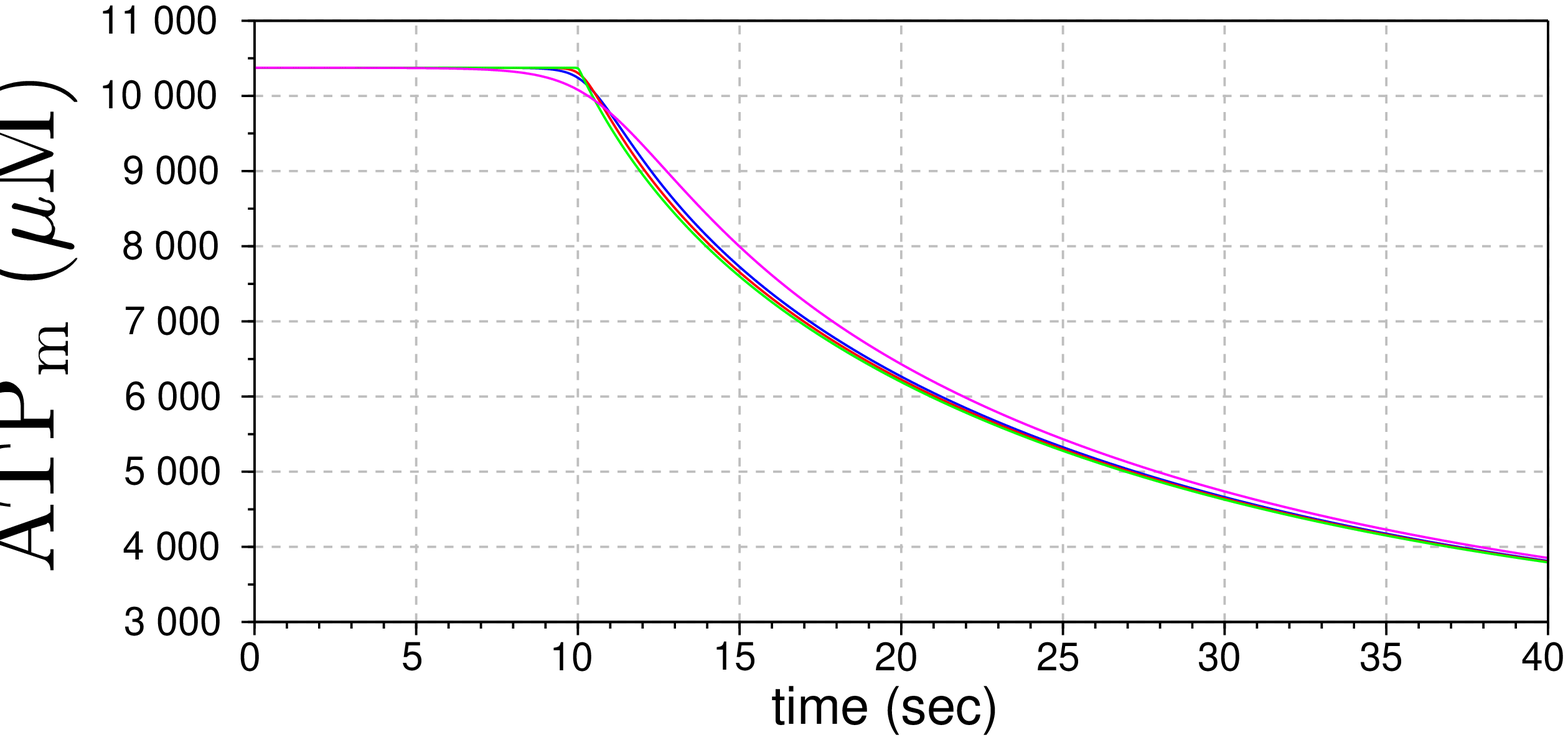}
  \includegraphics[width=0.5\linewidth]{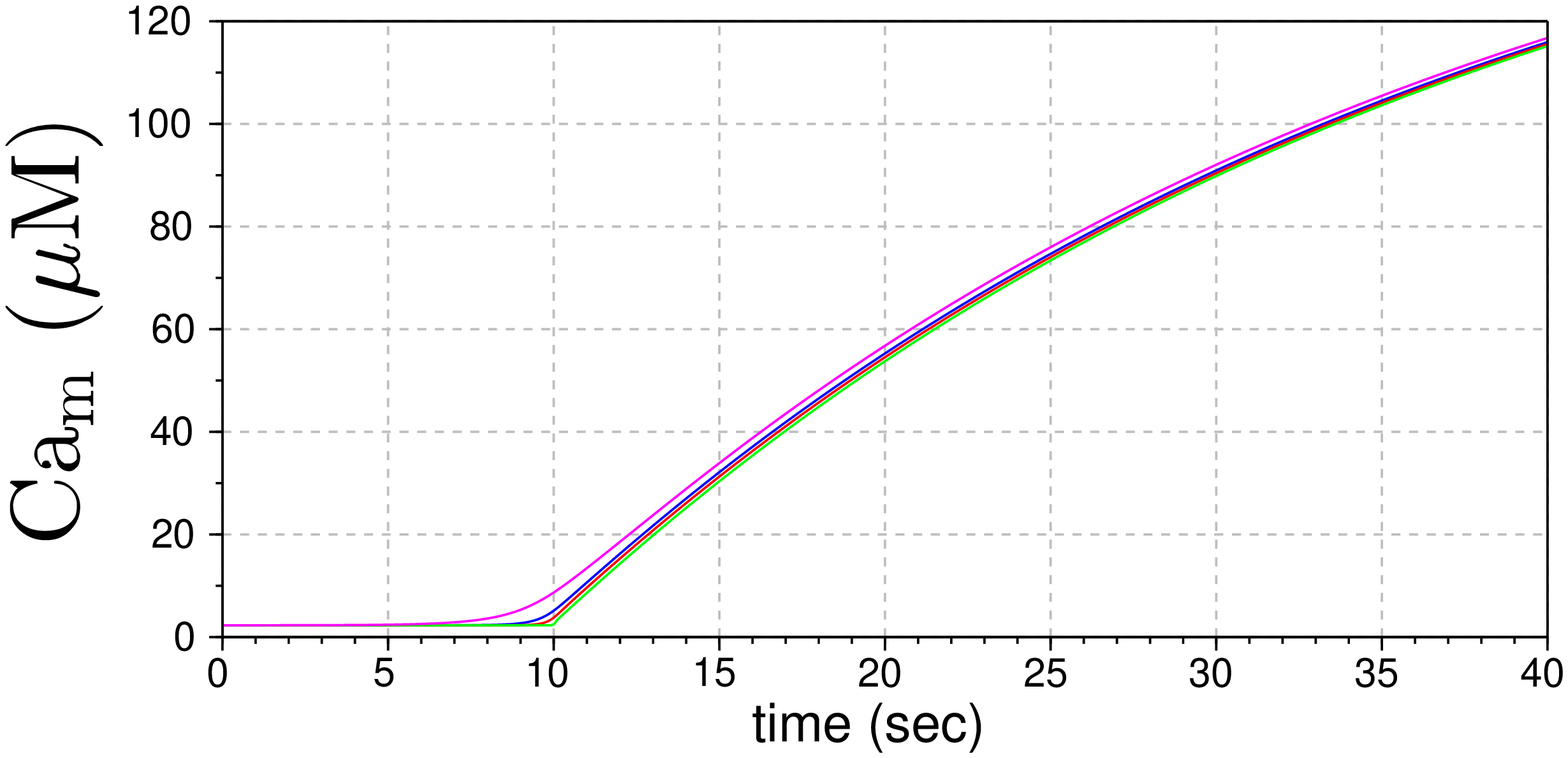}
  \includegraphics[width=0.5\linewidth]{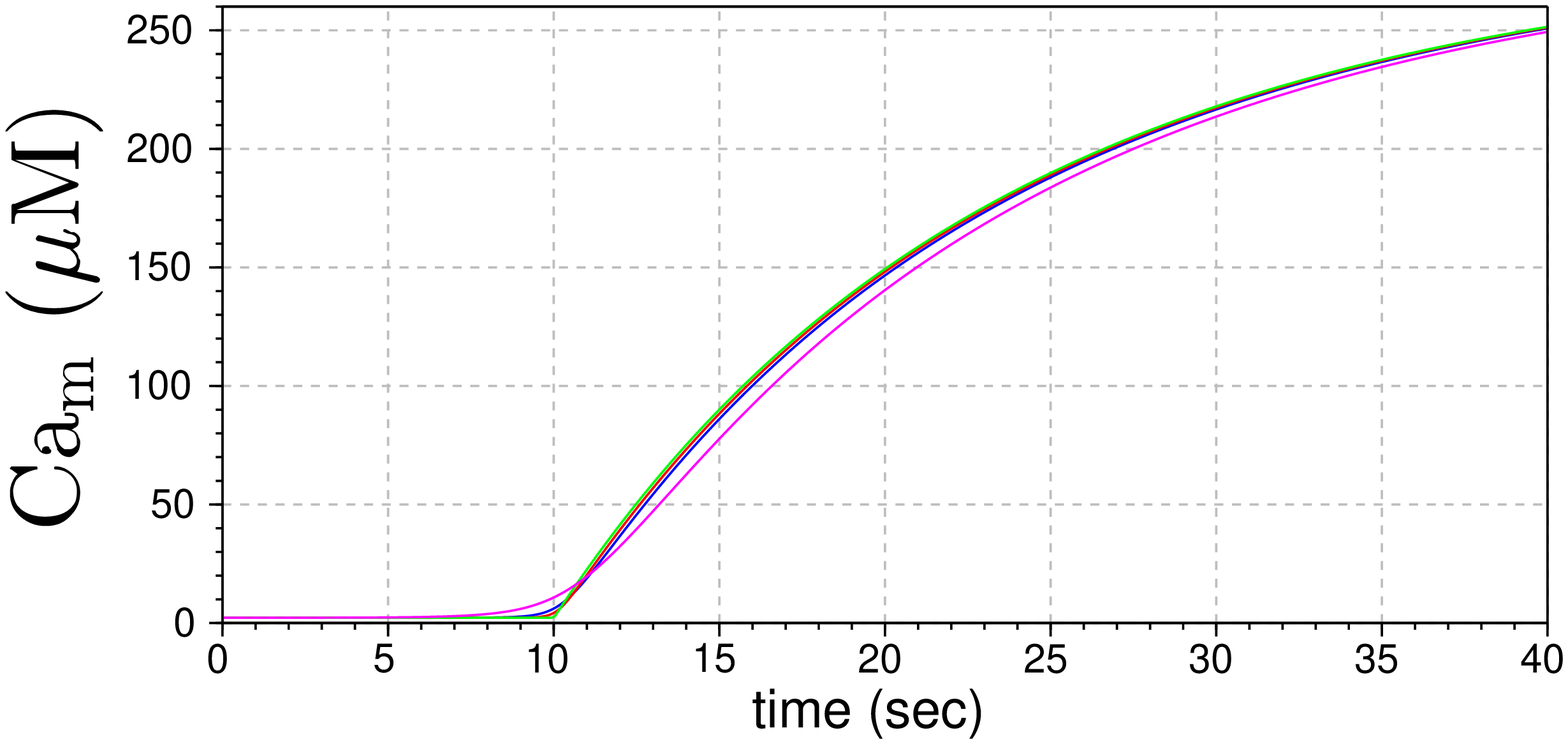}
  \includegraphics[width=0.5\linewidth]{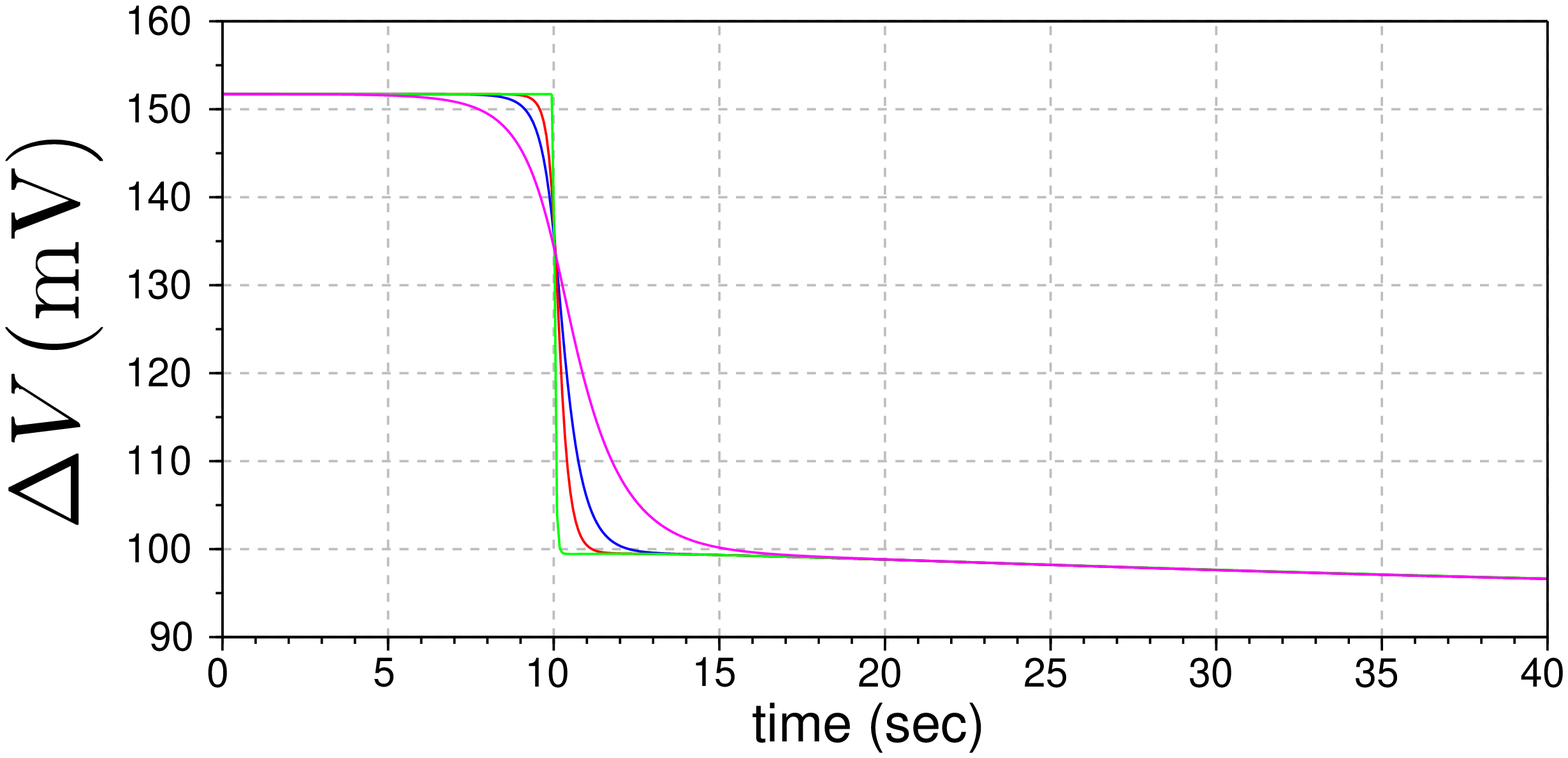}
  \includegraphics[width=0.5\linewidth]{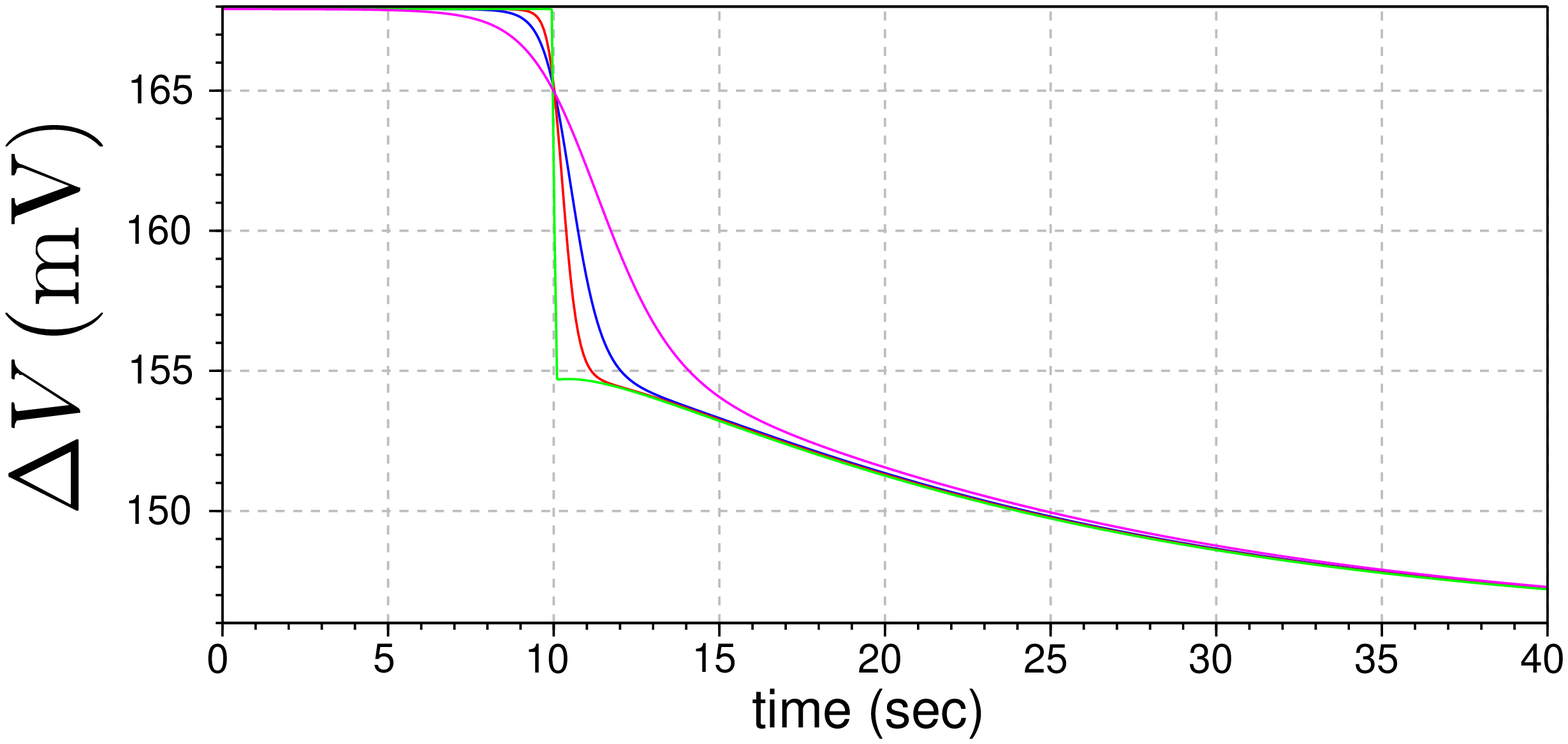}
\caption{Response of the equations (\ref{ee1})-(\ref{ee4}) to step-like inputs (\ref{tanh}). The  magenta, blue, red, and green curves correspond, respectively, to
$t_0= 2.5, 1, 0.5,$ and $0.02\,$seconds. 
The inertia of the system also increases considerably for higher values
of $v$, see the text for further details. The system is submitted to the input corresponding to the top graphics. The left column depicts the response for 
${\rm FBP} = 0.5\,\mu$M, while de right one corresponds to ${\rm FBP} = 10\,\mu$M.
See the text for further details.}
\label{fig6}
\end{figure}
This situation is analyzed and depicted in Fig. (\ref{fig6}). Besides the increasing of the relaxation times, for higher concentrations of FBP we observe a smoothing out
of the dynamical response of ${\rm NADH}_{\rm m}$. In particular, its overshooting  
present for large variations of ${\rm CA}_{\rm c}$ and low FBP disappears for the high FBP concentration case, compare Figures (\ref{fig5}) and (\ref{fig6}). By examining  this overshooting
in the dynamics of ${\rm NADH}_{\rm m}$, our rapidest variable, it is possible to estimate the critical time for which any adiabatic approximation should break. We can see from Figures (\ref{fig5}) and (\ref{fig6}) that the overshooting appears for transitions 
occurring in less than 2.5 seconds approximately. We do not expect any stationary response for input variables varying over periods smaller than this.

\section{Final Remarks}

We have revisited here the mathematical model for ATP production in mitochondria introduced recently by Bertram, Pedersen, Luciani, and Sherman (BPLS) in \cite{bertram} as a simplification of the more complete but intricate Magnus and Keizer's model
\cite{mk_1997,mk_1998_a,mk_1998_b}. We checked carefully all the approximations introduced in the BPLS model and found some inaccuracies for the approximations used for 
the  adenine nucleotide translocator rate 
$J_{\rm ANT}$ and for the calcium uniporter rate 
$J_{\rm uni}$. We proposed some enhanced approximations for such rates based on the original 
 Magnus and Keizer's model and analyzed some dynamical properties of the model. 
 Our results for the stationary regime indicate that  
the BPLS model is indeed globally stable, reinforcing its relevance to physiological quantitative studies, despite its simplicity when compared to the Magnus and Keizer's  model. We have considered also the non-stationary regime and detected a
effect which could be, in principle,  physiologically  interesting: the inertia of the system tends to increase considerably for high concentrations of cytosolic calcium and FBP, {\em i.e.},
  some response times of the model tend to increase considerably for high 
respiration inputs  
  ${\rm Ca}_{\rm c}$ and FBP. In particular, for 
  ${\rm Ca}_{\rm c}\approx 0.2\,\mu$M and  ${\rm FBP}\approx 0.5\,\mu$M, approximately
  10 seconds are necessary to ${\rm NADH}_{\rm m}$  and 
  $\Delta V$ attain homeostasis after a sudden increasing in ${\rm Ca}_{\rm c}$.
  The variables  ${\rm ATP}_{\rm m}$ and  ${\rm Ca}_{\rm m}$ are typically slower and need approximately 30 seconds to attain homeostasis in the same conditions. Keeping FBP constant and increasing ${\rm Ca}_{\rm c}$,  or keeping ${\rm Ca}_{\rm c}$ and
  increasing FBP, will imply a considerably increasing of this response time, {\em i.e.}, the system will take a longer time to attain homeostasis.

 It is interesting to notice that the dynamics of our enhanced model are qualitatively similar to the original BPLS one, despite the differences in the rates $J_{\rm ANT}$ and 
$J_{\rm uni}$ for physiological ranges, as depicted, for instance, in Fig. \ref{fig1}. This point can be understood from the fact that the value of $w_*$, which does not depends 
tightly on the details of such rates, 
 is almost constant and
corresponding to $\Delta V = 150\,$mV for   reasonable values of the inputs $v_*$ and $u_*$. For a fixed value of $\Delta V$, the numerical parameters in (\ref{JANT0})
can be fitted to provide a good adjustment for the real ATP dependence of (\ref{JANT1}). An inspection of Fig. 9 of \cite{bertram} reveals that the adjustment of their numerical parameters was probably checked for ${\rm ATP}\approx 3\,$mM and for $\Delta V \approx 160\,$mV, which is  close  to the physiological global fixed point (homeostasis), explaining why the asymptotic dynamics are not strongly affected by the inaccuracies in the BPLS approximations. On the other hand, we do not expect that the detected non-stationary effects be independent  on the details of $J_{\rm ANT}$ and
$J_{\rm uni}$. Such points certainly deserve further investigations.

\begin{acknowledgements}
The authors are grateful to FAPESP and CNPq for the financial support. AS wishes to thank Prof. Leon Brenig for the warm hospitality at the Brussels Free University, where part of this work was carried on.
\end{acknowledgements}


\begin{thebibliography}{99}

\bibitem{guyton} A. C. Guyton and J. E. Hall,  Textbook of Medical Physiology, Elsevier  (2006).

\bibitem{lehninger} D. L. Nelson and M. Cox,  Lehninger  Principles of Biochemistry, W. H. Freeman and Company (2004).

\bibitem{mk_1997}G. Magnus and J. Keizer, Am. J. Physiol. {\bf 273}, C717-C733 (1997).

\bibitem{mk_1998_a} G. Magnus and J. Keizer, Am. J. Physiol.  {\bf 274}, C1158-C1173 (1998).


\bibitem{mk_1998_b} G. Magnus and J. Keizer, Am. J. Physiol. {\bf 274}, C1174-C1184 (1998).

\bibitem{bertram} R. Bertram, M.G. Pedersen, D.S. Luciani, and A. Shermand, 
J. Theor. Biol. {\bf 243}, 575-586 (2006).

\bibitem{cortassa} S. Cortassa, M.A. Aon, E. Marban, R.L. Winslow, and B. O'Rourke,   Biophys. J. {\bf 84}, 2734-2755 (2003).

\bibitem{scilab} Scilab files are available at 
\href{http://vigo.ime.unicamp.br/atp}{\tt http://vigo.ime.unicamp.br/atp}


\end{thebibliography}
\end{document}